\documentclass[12pt]{article}

\usepackage[margin=1in]{geometry}  
\usepackage[singlespacing]{setspace}
\usepackage{epsfig}
\usepackage{graphics,graphicx,color}
\usepackage{amssymb,amsfonts,amsthm,amsmath}
\usepackage{authblk}
\usepackage{multirow}
\usepackage{mathrsfs}
\usepackage{enumerate}
\usepackage{natbib}
\usepackage{float}
\usepackage{algorithm}
\usepackage{algpseudocode}
\usepackage{subcaption} 
\usepackage{multirow}
\usepackage{url}
\usepackage[colorlinks=true, linkcolor=blue, citecolor=blue]{hyperref}
\usepackage{bm}
\usepackage{comment}

\newtheorem{theorem}{Theorem}
\newtheorem{example}{Example}[section]

\newcommand{\cov}{\mbox{cov}}

\DeclareMathOperator*{\E}{E}

\def\OLH{{\mbox{OLH}}}

\def\OA{{\mbox{OA}}}

\newcommand{\bX}{\mbox{\boldmath{$X$}}}
\newcommand{\bx}{\mbox{\boldmath{$x$}}}

\newcommand{\bD}{\mbox{\boldmath{$D$}}}

\newcommand{\bL}{\mbox{\boldmath{$L$}}}

\newcommand{\pkg}[1]{{\normalfont\fontseries{b}\selectfont #1}}
\let\proglang=\textit

\graphicspath{ {./Figures/} }

\title{Design of Experiments for Emulations: A Selective Review from a Modeling Perspective} 

\author[1]{Xinwei Deng}
\author[2]{Lulu Kang}
\author[3]{C. Devon Lin} 

\affil[1]{Department of Statistics, Virginia Tech}
\affil[2]{Department of Mathematics and Statistics, University of Massachusetts Amherst}
\affil[3]{Department of Mathematics and Statistics, Queen's University}
\begin{document}

\date{}
\maketitle

\begin{abstract}
Space-filling designs are crucial for efficient computer experiments, enabling accurate surrogate modeling and uncertainty quantification in many scientific and engineering applications, such as digital twin systems and cyber-physical systems.  
In this work, we will provide a comprehensive review of key design methodologies, including Maximin/miniMax designs, Latin hypercubes, and projection-based designs. Moreover, we will connect the space-filling design criteria, like the fill distance, to the Gaussian process performance. 
Numerical studies are conducted to investigate the practical trade-offs among various design types, with a discussion on emerging challenges in high-dimensional and constrained settings. 
The paper concludes with future directions in adaptive sampling and machine learning integration, providing guidance for improving computational experiments.

\end{abstract}

\section{Introduction}
In the era of digital transformation, \emph{Digital Twin} \citep{national2023foundational} has emerged as a powerful paradigm for simulating, predicting, and optimizing complex real-world systems through virtual models. 
By creating a dynamic, data-driven replica of physical systems, Digital Twin enables real-time monitoring, diagnostics, and decision-making. 
However, the fidelity and reliability of these virtual models heavily depend on uncertainty quantification (UQ) \citep{ghanem2017handbook, smith2024uncertainty}, a multidisciplinary field that bridges mathematics, statistics, and computational science. 
UQ seeks to characterize and mitigate uncertainties in model inputs, parameters, and outputs, ensuring robust predictions and actionable insights.

Computer experiments \citep{fang2002,santner2003design}, which replace costly or infeasible physical experiments with simulations, play a pivotal role in Digital Twin frameworks. 
These experiments allow researchers to explore system behaviors under controlled virtual conditions, making them indispensable in fields such as nuclear physics, climate modeling, and engineering design. 
Unlike physical experiments, computer simulations are cost-efficient, time-saving, and often the only viable option for studying extreme or hazardous scenarios. 
However, the accuracy of these simulations hinges on how effectively the input space is sampled. 
This is where the design of experiments (DoE) becomes critical.

A well-designed experiment ensures that the input space is thoroughly explored, enabling the surrogate model (e.g., Gaussian process regression) to capture complex, nonlinear relationships between inputs and outputs. Space-filling designs, such as Latin hypercubes, maximin distance designs, and low-discrepancy sequences, are particularly valuable because they distribute points evenly across the input domain, avoiding clustering and bias. These designs not only enhance prediction accuracy but also improve the robustness of uncertainty quantification.

This review synthesizes modern advancements in space-filling designs for computer experiments, focusing on their theoretical foundations, algorithmic implementations, and practical applications. We examine key criteria—such as fill distance, separation distance, and projection properties—that balance geometric uniformity with statistical optimality. Additionally, we discuss emerging challenges, including high-dimensionality, mixed variable types, and constrained design spaces, while highlighting innovative solutions like grouped orthogonal arrays and non-uniform space-filling designs. By integrating these methodologies, this work aims to guide researchers in selecting and constructing designs that maximize the efficacy of computer experiments for Digital Twin and other UQ-driven applications.

While several reviews on experimental design for computer simulations exist (e.g., \cite{levy2010computer}, \cite{pronzato2012design}, \cite{lin2015latin}, \cite{Joseph02012016}, \cite{GARUD201771}), this work fills critical gaps in three ways.
First, we cover the advances from the past decade—including methods for constrained spaces, multi-fidelity simulations, and mixed-variable inputs—that remain absent in earlier reviews. 
Second, with Digital Twin’s rising prominence, we offer newcomers a practical guide through comparative examples and tailored design recommendations for diverse problems. 
Finally, unlike intuitive discussions of space-filling properties in prior work, we ground justification in rigorous theory while maintaining accessibility, bridging classic results with modern insights.

The remainder of this work is organized as follows: Section 2 justifies the need for space-filling designs and connects them to surrogate modeling. Section 3 reviews classical and advanced space-filling designs, while Section 4 provides empirical comparisons. Finally, Section 5 concludes with future research directions.

\section{Space-Filling Design}

 A space-filling design is a method for selecting input variable settings to explore how responses depend on those inputs. By distributing points evenly across the entire input space, this approach ensures that the experimental region is well-represented by the design points.
In practice, when there is no prior preference or knowledge about the appropriate statistical model, a statistically robust strategy is to collect data from all regions of the design space. Space-filling designs enable this by supporting flexible statistical models and facilitating efficient exploration of the underlying response surface. Consequently, data collected using such designs—both inputs and outputs—provide a comprehensive understanding of the system’s input-output relationships. The concept of space-filling designs dates back to at least \cite{box1959basis}, who laid the foundation for response surface design selection. In computer experiments, the relationship between input factors and output responses is often highly complex and nonlinear. To ensure reliable and accurate emulation of the response surface, space-filling designs are essential — they strategically distribute design points across the input space, enabling comprehensive exploration of the system's behavior.

The concept of space-filling designs does not have a single, rigorous mathematical definition. Instead, various criteria have been proposed to measure how uniformly design points are distributed. Constructing optimal space-filling designs under these criteria is computationally and mathematically challenging — particularly in high-dimensional input spaces.  One widely used space-filling design is the Latin hypercube design \cite{mckay1979comparison}. A Latin hypercube of $n$ runs for $d$ input factors is represented by an $n \times d$ matrix, each column of which is a permutation of $n$ equally spaced levels. 
A key property of these designs is known as {\em one-dimensional uniformity} that when projected onto any individual dimension, the design points are evenly distributed across the variable’s range. Formally, a Latin hypercube design is defined as follows. For convenience, the $n$ levels are taken to be $1,2,\ldots,n$.  A Latin hypercube design, or a Latin hypercube sampling, $\bX$ in the design space $[0,1)^d$ can be generated using an $n \times d$ Latin hypercube  $\bL=(l_{ij})$ in the following way, 
\begin{eqnarray}\label{eq:dij}
x_{ij} = \frac{ l_{ij} - u_{ij} }{n}, \ \ \ i=1, \ldots, n, j = 1, \ldots, d,
\end{eqnarray}
\noindent where $x_{ij}$ is the $(i,j)$th entry of $\bX$ and $u_{ij}$'s are independent random numbers from $[0,1)$. The ``lattice sample'' due to \cite{patterson1954errors} corresponds to $\bL$ with $u_{ij} =0.5$ for all $(i,j)$'s.

\begin{figure}[!htb]
\centering
\includegraphics[width=0.7\textwidth]{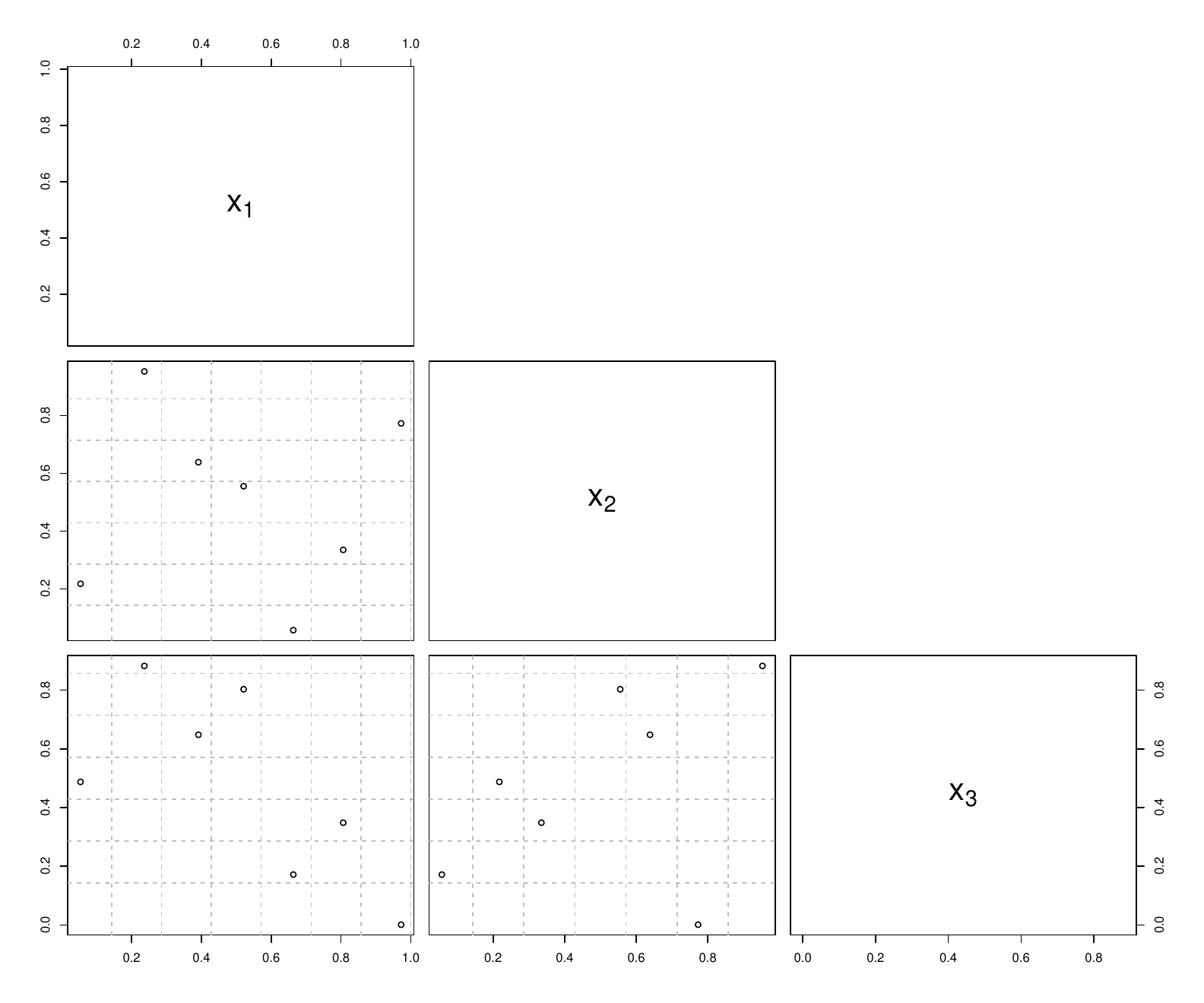}
\caption{The pairwise plot of the Latin hypercube design $\bX$ in Table~\ref{tab:lh1} for the three input variables  $X_1,X_2,X_3$}\label{fig:lhs1}
\end{figure}

Table~\ref{tab:lh1} displays a Latin hypercube $\bL$ of $7$ runs for $3$ input variables $X_1,X_2,X_3$ and a Latin hypercube design $\bX$ based on $\bL$. Figure~\ref{fig:lhs1} shows the pairwise plot of $\bX$ and illustrates the one-dimensional uniformity: when the seven points are projected onto each axis, there is exactly one point in each of the seven equally-spaced intervals.

\begin{table}[!htb]
\begin{center}
\caption{A $7 \times 3$ Latin hypercube $\bL$   and a Latin hypercube
design $\bX$ based on $\bL$}
\begin{tabular}{rrrrrrrr}
\multicolumn{3}{c}{$\bL$} & \multicolumn{2}{c}{} & \multicolumn{3}{c}{$\bX$}\\
    4  &  4  &  6& &  $\quad \quad $ &0.521  &0.555  &0.803\\
    5  &  1  &  2& &  $\quad \quad $ &0.663  &0.057  &0.172\\
    3  &  5  &  5& &  $\quad \quad $ &0.392  &0.638 & 0.648\\
    2  &  7  &  7& &  $\quad \quad $ &0.237  &0.953  &0.882\\
    1  &  2  &  4& &  $\quad \quad $ &0.054  &0.217  &0.487\\
    7  &  6  &  1& &  $\quad \quad $ &0.972 & 0.773  &0.001\\
    6  &  3  &  3& &  $\quad \quad $ & 0.806  &0.335  &0.348
\end{tabular}\label{tab:lh1}
\end{center}
\end{table}
 
The popularity of Latin hypercube designs stems from their theoretical justification for variance reduction in numerical integration. \cite{mckay1979comparison} demonstrated that when the integrand is monotonic in each input variable, the sample mean obtained from Latin hypercube designs has lower variance than that from simple random sampling. Explicit variance formulas were provided in the original work. Subsequent theoretical developments, including asymptotic normality and central limit theorems for Latin hypercube sampling, were established by \cite{stein1987large}, \cite{owen1992central}, and \cite{loh1996latin}.

A randomly generated Latin hypercube design  often exhibits poor space-filling characteristics. As illustrated in Figure~\ref{fig:lhs1} (particularly in the $X_2$ versus $X_3$ projection), such designs may display point clustering along diagonals, leaving substantial regions of the design space unexplored. This spatial clustering typically corresponds to high correlations among columns in the design matrix.  To overcome these limitations, researchers have developed enhanced Latin hypercube designs optimized using several key criteria:
\begin{itemize}
    \item \textbf{Distance-based}: Maximin and minimax; 
    \item \textbf{Orthogonality}: Minimizing column correlations \citep{owen1994controlling};
    \item \textbf{Projection properties}: Ensuring uniform coverage in lower-dimensional projections.
\end{itemize}
Construction methods for these improved designs have been extensively studied \citep{steinberg2006construction, lin2009construction, lin2010new, sun2017general}, with orthogonal and nearly-orthogonal Latin hypercubes being particularly important for many applications.

\section{Gaussian Process Emulator}

We first briefly review kriging in its popular form. Denote $\{\bm x_i, y_i\}_{i=1}^n$ as the $n$ pairs of input and output data from a certain computer experiment, and $\bm x_i\in \Omega \subseteq \mathbb{R}^d$ are the $i$th experimental input values and $y_i\in \mathbb{R}$ is the corresponding output. 
In this journal, we only consider the case of univariate response variable. 
To explain the need for space-filling designs, it is sufficient to focus on the simplest form of kriging, \emph{ordinary kriging}. 
It is built on the following model assumption of the response variable, 
\begin{equation}\label{eq:model}
y_i=\mu+ Z(\bm x_i)+\epsilon_i, \quad i=1,\ldots, n,
\end{equation}
where $\mu$ is either set to be 0 (when $\bm y$ is centered in pre-processing) or estimated with other unknown parameters. 
The random noise $\epsilon_i$'s are independently and identically distributed following $\mathcal{N}(0, \sigma^2)$. 
They are also independent of the other stochastic components of \eqref{eq:model}. 
We assume the GP prior on the stochastic function $Z(\bm x)$, which is denoted as $Z(\cdot)\sim GP(0, \tau^2 K)$, i.e., $\mathbb{E}[Z(\bm x)]=0$ and the covariance function
\[
\cov[Z(\bm x_1), Z(\bm x_2)]=\tau^2 K(\bm x_1, \bm x_2;\bm \theta).
\]
In most applications of computer experiments, we use the stationary assumption of $Z(\bm x)$, and thus the variance $\tau^2$ is a constant. 
The function $K(\cdot, \cdot;\bm \theta): \Omega \times \Omega \mapsto \mathbb{R}_{+}$ is the correlation of the stochastic process with hyperparameters $\bm \theta$. 
For it to be valid, $K(\cdot,\cdot;\bm \theta)$ must be a symmetric positive definite kernel function. 
Gaussian and Mat\'{e}rn kernel functions are among the most popular used ones and their definitions are
\begin{align*}
\text{Gaussian }& K(\bm x_1, \bm x_2; \bm \theta)=\exp\left\{-\sum_{j=1}^d \theta_j(x_{1j}-x_{2j})^2\right\}, \\
\text{Mat\'{e}rn }& K(\bm x_1, \bm x_2; \bm \theta, \nu)\propto \left(\nu \sum_{j=1}^d\theta_j|x_{1j}-x_{2j}|^2\right)^{\nu/2} B_{\nu}\left(2\left[\nu \sum_{j=1}^d\theta_j|x_{1j}-x_{2j}|^2\right]^{1/2}\right),
\end{align*}
with $\bm \theta \in \mathbb{R}^d$ and $\bm \theta\geq 0$. 
The two kernels are anisotropic in the sense that $\theta_j$'s are different for different dimensions. 
In isotropic kernels, or Radial Basis Function (RBF) kernels, $\theta_j$'s are the same for all dimensions. 

In terms of response $Y(\bm x)$, it follows a Gaussian process with the following mean and covariance, 
\begin{align*}
\mathbb{E}[Y(\bm x)]&=\mu, \quad \forall \bm x \in \Omega\\
\cov[Y(\bm x_1), Y(\bm x_2)]&=\tau^2 K(\bm x_1, \bm x_2;\bm \omega) + \sigma^2\delta(\bm x_1, \bm x_2), \quad \forall \bm x_1, \bm x_2\in \Omega,\\\
&=\tau^2\left[ K(\bm x_1, \bm x_2;\bm \omega) + \eta\delta(\bm x_1, \bm x_2)\right], 
\end{align*}
where $\delta(\bm x_1,\bm x_2)=1$ if $\bm x_1=\bm x_2$ and 0 otherwise, and $\eta=\sigma^2/\tau^2$. 
So $\eta$ is interpreted as the noise-to-signal ratio. 
For deterministic computer experiments, the noise component is not part of the model, i.e., $\sigma^2=0$ and $\eta=0$. 
However, a nugget effect, which is a small $\eta$ value, is usually included in the covariance function to avoid the singularity of the covariance matrix \citep{peng2014on}. 
The unknown parameter values of the GP model are $(\mu, \bm \theta, \tau^2, \eta)$. 

Following the empirical Bayes or maximum likelihood estimation, $\mu$ and $\tau^2$ have the tractable solutions
\begin{equation*}
\hat{\mu} =\frac{\bm y^\top (\bm K+\eta \bm I_n)^{-1}{\bf 1}}{{\bf 1}^\top (\bm K+\eta \bm I_n)^{-1}{\bf 1}}, \quad 
\hat{\tau}^2 =\frac{1}{n}(\bm y-\hat{\mu}{\bf 1})^\top (\bm K+\eta \bm I_n)^{-1} (\bm y-\hat{\mu}{\bf 1}),
\end{equation*}
where $\bm y$ is the vector of output observations, $\bm I_n$ is an identity matrix of size $n$, ${\bf 1}$ is a $n$-dim vector of 1's, $\bm K$ is the $n\times n$ kernel matrix whose entries are $\bm K_{ij}=K(\bm x_i, \bm x_j)$. 
Therefore, $\bm K$ is a symmetric positive definite matrix since $K$ is such a kernel. 
To estimate the remaining parameters $\eta$ and $\bm \theta$, we replace $\mu$ and $\tau^2$ by estimates in the likelihood, and maximize the updated likelihood, which is equivalent to solving
\[\min_{\eta,\bm \theta} n\log(\hat{\tau}^2)+\log\det(\bm K+\eta\bm I_n).\]
The prediction at any query point $\bm x$ conditioned on the observed data are the conditional mean
\[\hat{y}(\bm x)=\E(y(\bm x)|\bm y)=\hat{\mu}+\bm k(\bm x)^\top (\bm K+\eta \bm I_n)^{-1}(\bm y-\hat{\mu}{\bf 1})\]
or $\hat{y}(\bm x)=\bm k(\bm x)^\top (\bm K+\eta \bm I_n)^{-1}\bm y$ if we assume $\mu=0$. 
The vector $\bm k(\bm x)$ contains the correlation between $\bm x$ and $\bm x_i$'s, i.e., $\bm k(\bm x)^\top =[K(\bm x,\bm x_1),\ldots, K(\bm x, \bm x_n)]$. 

It turns out that the predictor $\hat{y}(\bm x)$, when $\mu=0$, is the same as the regularized kernel regression. 
For any symmetric positive definite kernel function defined on $\Omega\subset \mathbb{R}^d$, $K$ can induce a reproducing kernel Hilbert space, denoted by $\mathcal{H}_K$, which satisfies the two conditions: (1) for any $\bm x\in \Omega$, $K(\bm x, \cdot)\in \mathcal{H}_K$; and (2) for any $f\in \mathcal{H}_K$ and any $\bm x\in \Omega$, $\langle f,K(\bm x,\cdot)\rangle_{\mathcal{H}_K}=f(\bm x)$, also known as the reproducing property. 
Here, $\langle\cdot,\cdot\rangle_{\mathcal{H}_K}$ stands for the inner product of $\mathcal{H}_K$ that is deduced from a bilinear operator $<\cdot, \cdot>$. 
For any $f:\Omega \rightarrow \mathbb{R}$ and $f\in \mathcal{H}_K$, there exists such a set $\mathcal{X}=\{\bm x_1,\ldots,\bm x_n\}$ and $\bm c\in \mathbb{R}^n$ such that $f(\cdot)=\sum_{i=1}^n c_iK(\bm x_i,\cdot)$. 
For any $f, g\in \mathcal{H}_K$, which can be written into $f,g\in\mathcal{F}$ with $f=\sum_{i=1}^n c_iK(\bm x_i,\cdot)$ and $g=\sum_{j=1}^m d_jK(\bm z_j,\cdot)$, the inner product and the norm based on it are defined by 
\[\langle f,g\rangle_{\mathcal{H}_K}=\sum_{i=1}^n \sum_{j=i}^m c_id_j K(\bm x_i, \bm z_j), \quad ||f||_{\mathcal{H}_K}=\sum_{i,j=1}^n c_ic_j K(\bm x_i, \bm x_j)=\bm c^\top \bm K\bm c.\]
Given data $\{\bm x_i,y_i\}_{i=1}^n$, for any kernel function $K(\cdot,\cdot)$ and $f\in\mathcal{H}_K$, define the regularized loss function 
\begin{equation}\label{eq:Q}
Q_{\eta}(f,K,\mathcal{X},\bm y) = ||\bm y-\bm f||_2^2 + \eta \|f\|^2_{\mathcal{H}_K},
\end{equation}
where $\bm f$ is the vector $f(\bm x_i)$ for $i=1,\ldots,n$. 
Since $f\in \mathcal{H}_K$, we can write $f(\bm x_i)=\sum_{j=1}^n c_j K(\bm x_i, \bm x_j)$, and thus $\bm f=\bm c^\top \bm K$. 
Then \eqref{eq:Q} is equivalent to
\begin{equation*}
\min_{\bm c\in\mathbb{R}^n}Q_{\eta}(\bm c,K, \mathcal{X}, \bm y)=||\bm y-\bm c^\top \bm K||^2_2+\eta\bm c^\top \bm K\bm c,
\end{equation*}
which is quadratic function of $\bm c$. 
Therefore, the optimal solution is $\bm c^*=(\bm K+\eta \bm I_n)^{-1}\bm y$ and thus $\hat{f}(\bm x)=\hat{y}(\bm x)$ when $\mu=0$. 

\section{Justification of Space-Filling Designs}\label{sec:why}
In this section, we want to justify the two intuitions in a rigorous fashion and connect it with the popular surrogate model for computer experiments, Gaussian process (GP) regression, or \emph{kriging}. 

\subsection{Fill and Separation Distance}

In the field of function approximation, kriging, particularly when $\eta=0$, is also known as kernel function interpolation, or RBF interpolation when $K$ is a RBF kernel. 
There are many established theories on the error estimates of the interpolator $\hat{y}(\bm x)=\bm k(\bm x)^\top \bm K^{-1}\bm y$. 
Although such error bounds have slightly different versions, they all involve three parts, fill distance, the norm of the function $f$, and a constant independent of $\mathcal{X}$. 
Fill distance is defined by 
\begin{equation}\label{eq:h}
h_{\mathcal{X}, \Omega}=\sup_{\bm x\in \Omega} \min_{\bm x_j\in \mathcal{X}} ||\bm x-\bm x_j||_2.
\end{equation}
If $\Omega$ is a closed set, then $\sup_{\bm x\in \Omega}$ can be attained and should be changed to $\max_{\bm x\in \Omega}$. 
Using our notation, we recite the important result on the error bound of $f-\hat{y}$ from Chapter 11 of \cite{wendland2004scattered}.
\begin{theorem}\label{thm:error} \citep{wendland2004scattered}
 Suppose that $K\in C_{\nu}^k(\mathbb{R}^d)$ is conditionally positive definite of order $m$. Suppose further that $\Omega\subset \mathbb{R}^d$ is bounded and satisfies an interior cone condition. For $\bm \alpha \in \mathbb{N}_0^d$ with $\|\bm \alpha\|_1 \leq k/2$ and $\mathcal{X}=\{\bm x_1,\ldots, \bm x_n\} \subset \Omega$ satisfying $h_{\mathcal{X},\Omega} \leq h_0$ we have the error bound
 \[
 \|D^{\bm \alpha} f-D^{\bm \alpha} \hat{y}\|_{L_{\infty}(\Omega)} \leq C h_{\mathcal{X},\Omega}^{(k+\nu)/2-\|\bm \alpha\|_1}||f||_{\mathcal{H}_K}.
 \]
\end{theorem}
\emph{Remark:}
There are several technical definitions that might be less familiar to readers. 
The notation $\bm \alpha$ is derivative order for the $d$-dimensions and $D^{\bm \alpha}$ denotes the derivative operator of the orders and dimensions specified by $\bm \alpha$. 
A conditionally positive definite kernel is a function that, when used to create the Gram matrix (such as the kernel matrix in kriging), results in a matrix that is conditionally positive definite, meaning it is positive semi-definite when the sum of the coefficients is zero. 
The order $m$ refers to the highest degree of polynomials that can be reproduced by the kernel associated with the conditionally positive definite kernel.
Thus, conditionally positive definite kernels include all positive definite kernels, such as the Gaussian and Mat\'{e}rn kernels we have showed earlier. 
The set $C_{\nu}^k(\mathbb{R}^d)$ is H\"{o}lder space consisting of all functions $f\in C^k(\mathbb{R}^d)$ (continuous to the $k$th degree) whose derivatives of order $k$ satisfy $D^{\bm \alpha}f(\bm x)=O(\|\bm x\|_2^{\nu})$ for $\|\bm x\|_2^{\nu}\to 0$. 
If $\Omega$ is bounded and satisfies an interior cone condition, simply put, it implies if $\bm x\in \Omega$, a cone with $\bm x$ as vertex is also in $\Omega$.
Here $C$ and $h_0$ are both some constant that only depend on $\Omega$. 
Theorem \ref{thm:error} is general and covers the error bound of the derivatives with order $\|\bm \alpha\|_1\leq k/2$ of $f-\hat{y}$ for a wide range of kernel functions. 
As a special case, 
\[
||f-\hat{y}||_{L_{\infty}(\Omega)}\leq C h_{\mathcal{X},\Omega}^{(k+\nu)/2}||f||_{\mathcal{H}_K}.
\]
Based on the error bound, we can see that if the fill distance $h_{\mathcal{X},\Omega}\to 0$, the largest possible error between the underlying function $f$ and the interpolator converges to 0 as well. 
For any given $\Omega$, the fill distance only depends on the design $\mathcal{X}$. 
Based on this result, if kriging is to be used as the surrogate model for a computer experiment, a design that leads to small fill distance is more suitable. 

Separation distance, which is defined as 
\[
h_{\mathcal{X},\text{sep}}=\min_{\bm x,\bm x'\in \mathcal{X}} ||\bm x-\bm x'||_2, 
\]
is also related to the accuracy of kriging model, but it was not discussed in the field function approximation. 
Regarding the difference and relationship of separation and fill distance, \cite{tuo2020kriging} derived some important results, which were further explained by \cite{he2024efficient}. 
The results said, based on isotropic Mat\'{e}rn covariance function, if the estimated roughness parameter of the kernel $\hat{\nu}$ is less than or equal to the true roughness parameter value $\nu_0$, i.e., 
$\hat{\nu}\leq \nu_0$, there exist constants $h_0\in (0, 1]$, $C_1, C_2>0$ and $C_3>e$, such that for any $\mathcal{X}$, $h_{\mathcal{X}, \Omega}\leq h_0$ and any $t>0$, with probability at least $1-\exp[-t^2/(C_1h_{\mathcal{X},\Omega}^{2\hat{\nu}})]^2$,  
\[
\sup_{\bm x\in \Omega}|y(\bm x)-\hat{y}(\bm x)|\leq C_2 \tau h_{\mathcal{X}, \Omega}\log^{1/2}(C_3/h_{\mathcal{X},\Omega})+t.
\]
Here $y(\bm x)$ is the Gaussian process in the model assumption with $\mu=0$ and $\eta=0$. 
We can see that this error bound only depends on the fill distance, which agrees with the same above conclusion from \cite{wendland2004scattered}, even though the this error model is with respect to the GP assumption and different from the typical one for function approximation.  
However, if $\hat{\nu}>\nu_0$, there exist constants $h_0\in (0, 1]$, $C_1, C_2>0$ and $C_3>e$, such that for any $\mathcal{X}$, $h_{\mathcal{X}, \Omega}\leq h_0$ and any $t>0$, with probability at least 
$1-\exp[-t^2/(C_1\tau^2h_{\mathcal{X},\Omega}^{2\hat{\nu}}\left(h_{\mathcal{X},\Omega}/(h_{\mathcal{X},\text{sep}}/2)\right)^{2(\hat{\nu}-\nu_0)})]$,  
\[
\sup_{\bm x\in \Omega}|y(\bm x)-\hat{y}(\bm x)|\leq C_2 \tau h_{\mathcal{X}, \Omega}^{\hat{\nu}}\left(h_{\mathcal{X},\Omega}/(h_{\mathcal{X},\text{sep}}/2)\right)^{\hat{\nu}-\nu_0}\log^{1/2}(C_3/h_{\mathcal{X},\Omega})+t.
\]
More precisely, it is increasing with respect to $h_{\mathcal{X},\Omega}$ and the ratio $h_{\mathcal{X},\Omega}/h_{\mathcal{X}, \text{sep}}$. 
Therefore, to reduce the error bound, we need to simultaneously reduce the fill distance and increase the separation distance. 
This point was also advocated by \cite{chen2024selecting}. 

A space-filling design, which aims to spread the design points evenly in the design space, is directly or indirectly minimizing the fill distance or maximizing the separation distance. 
For example, the Maximin space-filling design of size $n$ \citep{johnson1990minimax} is the solution of 
\[\max_{\mathcal{X}\subset \Omega}\min_{\bm x, \bm x'\in \mathcal{X}} ||\bm x-\bm x'||_2, \]
which directly maximizes the separation distance and pushes the design points evenly distributed in $\Omega$.
On the other hand, miniMax space-filling design of size $n$ \citep{johnson1990minimax} is the solution of 
\[
\min_{\mathcal{X}\subset \Omega}\max_{\bm z\in \Omega} \text{dist}(\bm z, \mathcal{X}), 
\]
where $\text{dist}(\bm z, \mathcal{X})=\min_{\bm x\in \mathcal{X}} ||\bm z-\bm x||_2$. 
\cite{pronzato2017minimax} further explained that 
\[\max_{\bm z\in \Omega} \text{dist}(\bm z, \mathcal{X})=\max_{\bm x\in \Omega}\min_{i=1,\ldots,n}||\bm x-\bm x_i||_2, \]
which is a direct approximation of the fill distance in \eqref{eq:h}. 
Therefore, miniMax design aims to minimize the fill distance.

Based on the discussion, miniMax is a more appealing space-filling criterion, because Maximin design alone is not enough to control the upper bound on the approximation error of kriging. 
However, the optimization problem of the miniMax design is much more challenging than the Maximin design. 
More details of the types of designs are in Section \ref{sec:design}. 

\section{Designs with Space-Filling Properties}\label{sec:design}

This section reviews several widely-used classes of space-filling designs. We begin with Latin hypercube designs, covering three important variants including distance-based Latin hypercube designs, orthogonal and nearly orthogonal Latin hypercube designs and Projection-based Latin hypercube designs. 
 We then examine space-filling designs that extend beyond the Latin hypercube framework, including designs for constrained experimental regions and specialized designs for advanced computer experiments.

\subsection{Distance-based Latin Hypercube Designs}\label{sec:distancelhd}

A natural approach to improving the space-filling properties of Latin hypercube designs is to optimize them using distance-based criteria. Before introducing these criteria, we first define some notations. Let $\bm X$ denote a design matrix.  Define the $L_q$-distance between two runs $\bm x_i$ and $\bm x_j$ of $\bm X$ as:
\begin{equation}\label{eq: dist-eulician}
    d_q(\bm x_i, \bm x_j) = \left(\sum_{k=1}^{m} |x_{ik} - x_{jk}|^q \right)^{1/q},
\end{equation}
where $q$ is an integer with $q=1$ and $q=2$ corresponding to  the rectangular and Euclidean distances, respectively. Note that the distance in \eqref{eq: dist-eulician} is just a commonly used  distance for the continuous variables. One can also consider other types of distance.  Let $\mathcal{X}$ be the design space throughout. 

{\bf Maximin Distance Criterion.} 
Based on the $L_q$-distance, one can define the $L_q$-distance of design $\bm X$ as:
\begin{equation}\label{eq:dq}
    d_q(\bm X) = \min\{ d_q(\bm x_i, \bm x_j), 1 \leq i < j \leq n \}.
\end{equation}
A design $\bm X$ is called a maximin $L_q$-distance design if it has the largest $d_q(\bm X)$ value. 
That is
\begin{align*}
\bm X_{Mm} = \arg \max_{\bm X \in  \mathcal{X}} \min_{i, j} d_q(\bm x_i, \bm x_j).
\end{align*}
This criterion attempts to place the design points such that no two points are too close to each other. 

{\bf Minimax Distance Criterion.} 
A slightly different approach to achieving a well-distributed design is to ensure that every point in the design space $\mathcal{X}$ is close to at least one design point in $\bX$. This leads to the {\em minimax distance} criterion, which seeks a design $\bX$ of $n$ points in $\mathcal{X}$ that minimizes the maximum distance between any arbitrary point $\bx \in \mathcal{X}$ and the closest design point in $\bX$. Formally, this criterion minimizes  
\begin{equation*}
    \max_{\bx \in \mathcal{X}} d_q(\bx, \bX),
\end{equation*}
where $d_q(\bx, \bX)$ denotes the $L_q$ distance between $\bx$ and its nearest neighbor in $\bX$, defined as  
\begin{equation*}
    d_q(\bx, \bX) = \min_{\bx_i \in \bX} d_q(\bx, \bx_i),
\end{equation*}
with $d_q(\bx, \bx_i)$ given in (\ref{eq: dist-eulician}) for any specified value of $q$.

Alternative distance measures have been explored in the literature; see, for example, \cite{audze1977new} and \cite{MoonDeanSantner2011}.
Although minimax designs  have many useful
applications and enjoy desirable space-filling properties,  
there has been little work in developing algorithms for generating these designs, due to its computational
complexity. Notable exceptions are  \cite{van2008two}, \cite{he2017interleaved} and \cite{mak2018minimax}. 
We now focus on the maximin distance criterion. A maximin design is asymptotically D-optimal under a Gaussian process model as the correlations become weak. Consequently, it is also asymptotically optimal under the maximum entropy criterion \citep{shewry1987maximum}.

To express the maximin distance criterion in a more convenient way, \cite{morris1995} proposed the $\phi_p$ criterion which is defined as:
\begin{equation}\label{eq:phi-p criteria}
    \phi_p = \left( \sum_{i=1}^{n-1} \sum_{j=i+1}^{n} d_q(\bx_i, \bx_j)^{-p} \right)^{1/p}.
\end{equation}
The $\phi_p$ criterion in \eqref{eq:phi-p criteria} is asymptotically equivalent to the maximin distance criterion as $p \to \infty$.  The value of 
$p$ is chosen based on the design size, typically ranging from 5 for small designs to 20 for moderate-sized designs and 50 for large designs.

Maximin designs are likely to have clumped projections onto one-dimension.  Thus, such designs may not possess desirable one-dimensional uniformity which is guaranteed by Latin hypercube designs. To strike the balance,  \cite{morris1995} examined maximin designs within Latin hypercube designs.    Both algorithmic and algebraic methods have been employed to construct maximin Latin hypercube designs. For examples of algebraic constructions, see \cite{lin2016general}, \cite{xiao2017construction}, \cite{li2021method}, \cite{yin2023construction} and \cite{yuan2025construction}, while \cite{lin2015latin} provides a summary of algorithmic approaches. To the best of our knowledge, the \pkg{SLHD} package in \proglang{R} by \cite{ba2015optimal} implements the most efficient current algorithm, as revealed in Example~\ref{eq:designs}. Few work on algebraic construction provides publicly accessible and reproducible codes. Beyond the construction of maximin Latin hypercubes, the literature also explores the theoretical properties of the maximin distance designs. For example, \cite{zhou2015space} derived the upper bounds of $d_q(\bm X)$ in (\ref{eq:dq}). 

Although this idea of maximin Latin hypercube designs sounds simple,
 finding them can be computationally intensive and time-consuming, particularly for larger values of 
$n$ and  $d$. As a result, most studies in the literature focus on heuristic maximin Latin hypercube designs, where `heuristic' implies that the separation distance is not guaranteed to be maximal. However, for two-dimensional maximin Latin hypercube designs with
$q=1$, \cite{van2007maximin} have obtained `optimal' solutions. In addition, they used a branch-and-bound algorithm to construct two-dimensional maximin Latin hypercube designs with $q=2$ for  $n \leq 70$. These designs were previously available online via the website \url{https://spacefillingdesigns.nl}; however, the website is no longer accessible. 
  
\begin{example}\label{eq:designs}
We assess the performance of various algorithmic methods for generating maximin Latin hypercube designs. The methods compared include: the maximinSLHD function from the \pkg{SLHD} package in \proglang{R} \citep{SLHD},  the maximinESE\_LHS and maximinSA\_LHS functions from the \pkg{DiceDesign} package in \proglang{R}\citep{DiceDesign}. Although the \pkg{maximin} package in \proglang{R} \citep{maximin} can construct maximin Latin hypercube designs, it fails to produce the designs for six and more input variables due to the storage issue, and thus we do not include it in comparison. We consider run sizes $n$ of (30, 50, 70, 100, 150, 200) and the number of input variables $d$ of (3, 6, 9, 12, 15). For each $(n, d)$ combination and each method, we generate 100 designs and evaluate them using the criteria in   (\ref{eq:phi-p criteria}) and (\ref{eq:dq}), with $q = 2$ and $p = 50$.  Figure~\ref{fig:dist} presents the best performance for each method and case, reporting the minimum of 100 values for (\ref{eq:phi-p criteria}) and the maximum of 100 values for (\ref{eq:dq}). The figure shows that among the three algorithmic approaches, the \pkg{SLHD} package in \proglang{R} consistently produced maximin Latin hypercube designs with the smallest values of (\ref{eq:phi-p criteria}) and the largest values of (\ref{eq:dq}) across all sample sizes and the number of input variables  considered. Therefore, the \pkg{SLHD} package in \proglang{R} demonstrates the best performance in terms of the maximin distance criteria.

\begin{figure}[!htb]
     \centering
     \begin{subfigure}[b]{0.48\textwidth}
         \centering
         \includegraphics[width=\textwidth]{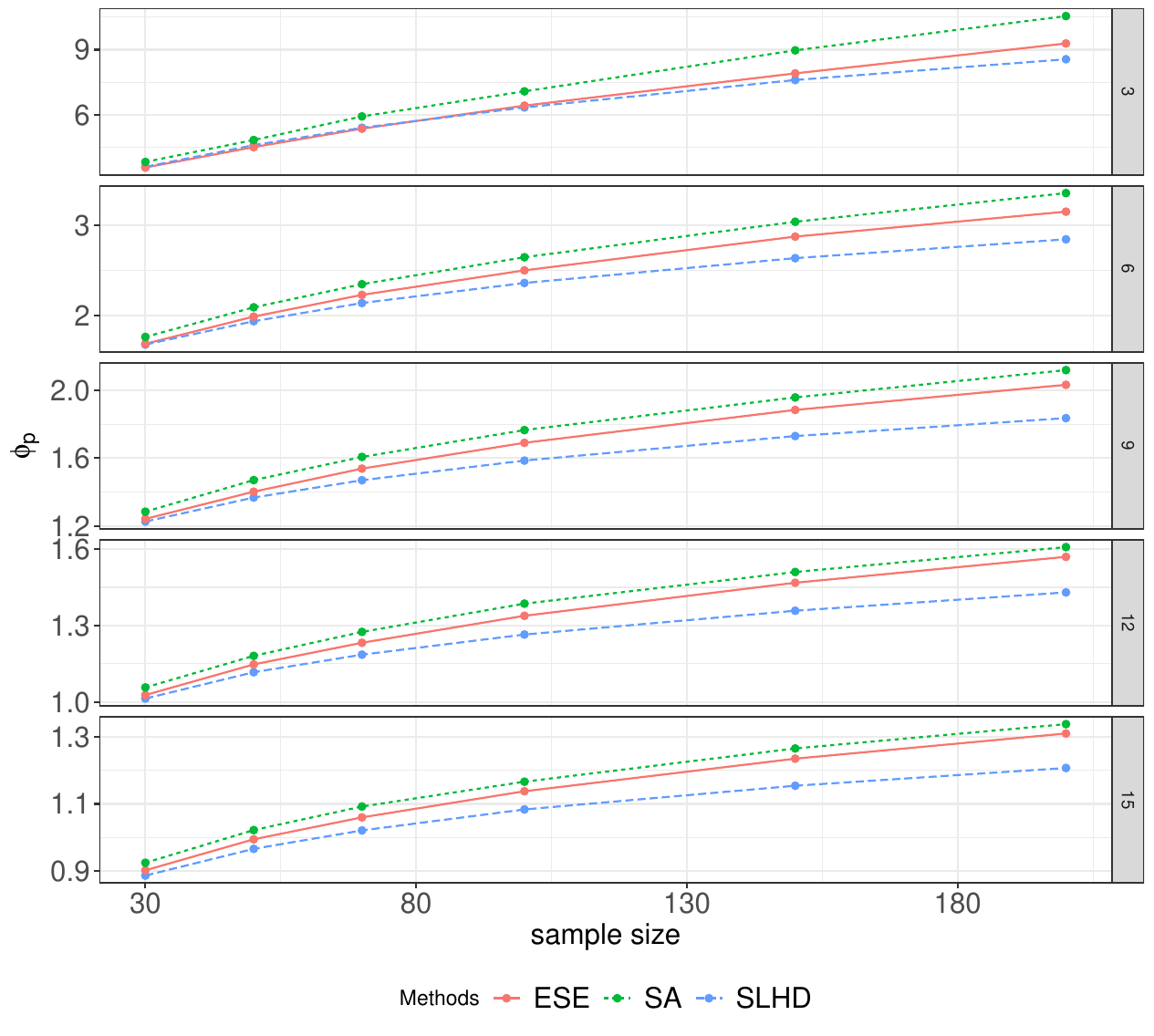}
         \caption{Criterion in (\ref{eq:phi-p criteria})}
         \label{fig:phip}
     \end{subfigure}
     \begin{subfigure}[b]{0.48\textwidth}
         \centering
         \includegraphics[width=\textwidth]{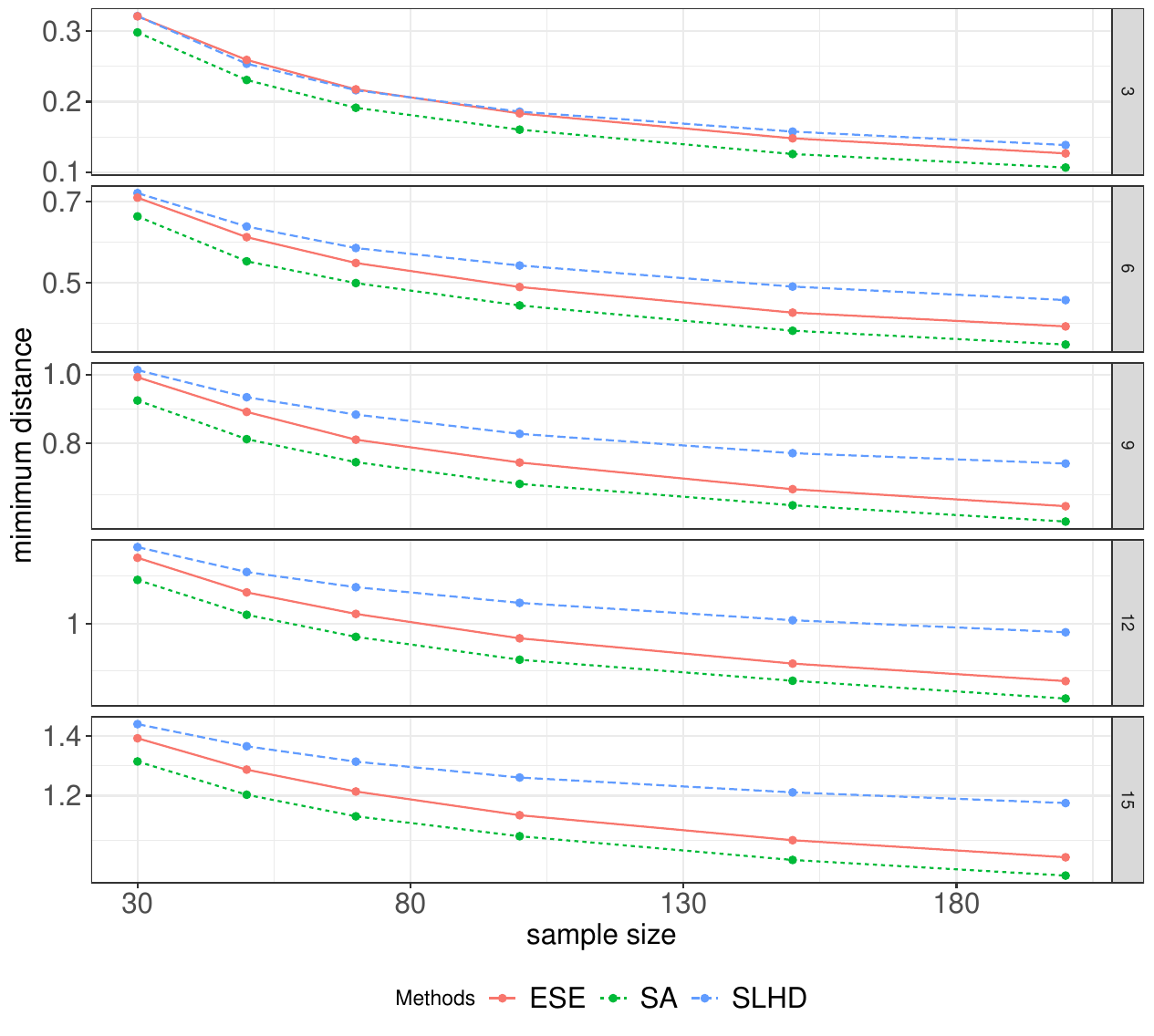}
         \caption{Criterion in (\ref{eq:dq}) }
         \label{fig:minidist}
     \end{subfigure}   
\caption{The performance in terms of (\ref{eq:phi-p criteria}) and  (\ref{eq:dq}) of maximin Latin hypercube designs generated by the $maximinSLHD, maximinESE\_LHS,maximinSA\_LHS $ functions}\label{fig:dist}
\end{figure}

\end{example}

\subsection{Orthogonal and Nearly Orthogonal Latin Hypercube Designs}\label{sec:olhd}

This section discusses the properties and constructions of Latin hypercube designs   with zero or minimal column-wise correlations across all two-dimensional projections. Such designs are referred to as \emph{orthogonal} and \emph{nearly orthogonal} Latin hypercube designs.

Orthogonal Latin hypercube designs are particularly valuable for fitting data using main effects models, as they enable uncorrelated estimates of linear main effects. Another motivation for pursuing orthogonal or nearly orthogonal Latin hypercube designs stems from their relationship with space-filling designs: while these designs may not necessarily be space-filling themselves, optimal space-filling designs should exhibit orthogonality or near-orthogonality.  This relationship suggests a strategy for identifying space-filling designs by searching within the class of orthogonal or nearly orthogonal Latin hypercube designs. Other theoretical and practical justifications for these designs can be found in \cite{iman1982distribution}, \cite{owen1994controlling}, \cite{tang1998selecting}, \cite{joseph2008orthogonal}, among others.

Extensive research has been conducted on the construction of orthogonal and nearly orthogonal Latin hypercube designs. The foundation for this research was established by \cite{ye1998orthogonal}, who pioneered the construction of orthogonal Latin hypercubes with $n = 2^m$ or $2^m + 1$ runs and $k = 2m - 2$ factors, where $m \geq 2$. Subsequent work has expanded upon these initial results in several directions: \cite{cioppa2007efficient} extended Ye's construction to accommodate more columns for given run sizes;  \cite{steinberg2006construction} developed orthogonal Latin hypercube designs with run sizes $n = 2^{2^m}$ by rotating groups of factors in two-level regular fractional factorial designs; \cite{pang2009construction} generalized this approach to $p^{2^m}$ runs (where $p$ is prime) with up to $(p^{2^m}-1)/(p-1)$ factors using $p$-level regular factorial designs; \cite{lin2008} introduced an algorithmic approach for small run sizes ($n \leq 20$) by sequentially adding columns to existing designs. Recent advances have focused on creating orthogonal Latin hypercube designs with more flexible run sizes and higher factor-to-run-size ratios. Of particular note are the general and computationally efficient methods proposed by \cite{lin2010new} and \cite{sun2017general}, which encompass the results presented in Table~\ref{table:large}.  Additional construction methods can be found in the works of \cite{lin2009construction}, \cite{sun2009construction}, \cite{georgiou2009orthogonal}, \cite{sun2010construction}, \cite{yangconstruction}, and \cite{sun2017method}, demonstrating the continued development and refinement of orthogonal Latin hypercube design construction methods.

A fundamental theoretical question in the study of orthogonal Latin hypercube designs  is determining the maximum number of columns, denoted $k^*$, that can be constructed for a given run size $n$ while maintaining orthogonality.  Theorem~\ref{theo:exist} establishes the following key results: $k^* = 1$ when $n = 3$ or $n = 4m + 2$ for any non-negative integer $m$; and $\geq 2$ for all other cases.  \cite{lin2010new} significantly strengthened these theoretical bounds by deriving more precise limitations on $k^*$ across various run sizes. Their work provides deeper insight into the structural constraints governing orthogonal Latin hypercube constructions.

\begin{theorem}\label{theo:exist}
There exists an orthogonal Latin hypercube of $n \geq 4$ runs with
more than one factor if and only if $n \neq 4m + 2$ where
$m$ is an integer.
\end{theorem}

\cite{lin2010new} made substantial theoretical advances by establishing tight upper bounds for $k^*$ through an innovative combinatorial framework. Their analysis revealed that:  exact maximum column counts $k^*$ for specific classes of run sizes $n$; improved lower bounds through constructive proofs for general cases the fundamental trade-off between orthogonality and design dimensionality. This work not only refined the existence conditions from Theorem~\ref{theo:exist} but also characterized the geometric and algebraic constraints underlying orthogonal Latin hypercube structures. The results provide both a theoretical benchmark and practical guidance for design construction.

\begin{theorem}\label{prop:m}
The maximum number $k^*$ of factors for an orthogonal
Latin hypercube of  $n = 16m + j$ runs
has a lower bound given below: \\
(i) $k^* \geq 6$ for all $n=16m+j$ where $m \geq 1$ and
$j \neq 2, 6, 10, 14$; \\
(ii) $k^* \geq 7$ for $n=16m + 11$ where $m \geq 0$; \\
(iii) $k^* \geq 12$ for $n=16m, 16m + 1$ where $m \geq 2$; \\
(iv) $k^* \geq 24$ for $n=32m, 32m + 1$ where $m \geq 2$; \\
(v) $k^* \geq 48$ for $n=64m, 64m + 1$ where $m \geq 2$.
\end{theorem}

The preceding theorem establishes a general lower bound for the maximum number of factors $k^*$ in an orthogonal Latin hypercube design with $n$ runs. We now present a comprehensive summary of the best currently known lower bounds for $k^*$ in $\OLH(n,k^*)$ designs across all existing construction methods, covering run sizes $n \leq 512$, where $\OLH(n,k)$
to denote an orthogonal Latin hypercube of $n$ runs for $k$ factors. Table~\ref{table:small} enumerates the optimal lower bounds for $k^*$ in the small run size regime. Notably: For all cases except $n=16$, the bounds were achieved through the algorithmic approach of \cite{lin2008}; The $n=16$ case yields a superior bound with $k^*=12$, as demonstrated by \cite{steinberg2006construction} using rotational methods. 
Table~\ref{table:large} presents the current best lower bounds for $k^*$ in this range and the corresponding construction methods achieving these bounds.  This systematic compilation enables direct comparison of methodological effectiveness while identifying opportunities for further improvement in orthogonal Latin hypercube construction.
 
\begin{table}[t]
\caption{The best lower bound $k$ on the maximum number $k^*$ of
factors in $\OLH(n,k^*)$ for $n \leq 24$ }
\begin{center}
\scalebox{0.9}{
\begin{tabular}{l|cccccccccccccccc}
\hline \hline
$n$ & 4 & 5 & 7 & 8 & 9 & 11 & 12 & 13 & 15 & 16 & 17 & 19 & 20 & 21 & 23 & 24 \\
\hline $k$ & 2 & 2 & 3 & 4 & 5 & 7 & 6 & 6 & 6 & 12 & 6 & 6 & 6& 6 & 6 & 6\\
\hline
\end{tabular}}
\end{center}\label{table:small}
\end{table}

\begin{table}[t]
\caption{The best lower bound $k$ on the maximum number $k^*$ of
factors in $\OLH(n,k^*)$ for $512 
\ge n>24$ }
\begin{center}
\scalebox{0.8}{
\begin{tabular}{l l lll ccccccc lllll}
\hline \hline
$n$&&$k$& &Reference             & & & & & & &  $n$&& $k$&& Reference    \\
\hline
25 & &12& &\cite{lin2009construction}& & & & & & &  145&& 12 && \cite{lin2010new}  \\
27 & &12& &\cite{sun2017general}& & & & & & &  160&& 24 && \cite{lin2010new} \\
32 & &24 & &\cite{sun2017general} & & & & & & &   161&& 24 && \cite{lin2010new} \\
33 & &16 & &\cite{sun2009construction}& & & & & & &  169&& 84 &&  \cite{lin2009construction} \\
48 & &12 & &\cite{lin2010new}     & & & & & & & 176&& 12 &&\cite{lin2010new}\\
49 & &24 & &\cite{lin2009construction}       & & & & & & & 177&& 12 &&\cite{lin2010new}\\
64 & &48 & &\cite{sun2017general} & & & & & & &  192&& 48 && \cite{lin2010new} \\
65 & &32 & &\cite{sun2009construction} & & & & & & &  193&& 48 && \cite{lin2010new}\\
80 & &12 & &\cite{lin2010new}    & & & & & & &  208&& 12 && \cite{lin2010new} \\
81 & &50 & &\cite{lin2009construction}       & & & & & & & 209&& 12 && \cite{lin2010new} \\
96 & &24 & &\cite{lin2010new}       & & & & & & & 224&& 24 && \cite{lin2010new} \\
97 & &24 & &\cite{lin2010new}    & & & & & & &  225&& 24 && \cite{lin2010new} \\
112& &12 & &\cite{lin2010new}      & & & & & & &  240&& 12 && \cite{lin2010new} \\
113& &12 & &\cite{lin2010new}     & & & & & & &  241&& 12 && \cite{lin2010new}  \\
121& &84 & &\cite{lin2009construction}       & & & & & & &  243&& 80 && \cite{sun2017general} \\
125& &58 & &\cite{sun2017general}       & & & & & & &  256&& 248&&\cite{steinberg2006construction}\\
128& &96 & &\cite{sun2017general} & & & & & & &   343 && 168 &&\cite{sun2017general}  \\
129& &64 & &\cite{sun2009construction} & & & & & & &  512&& 496 &&\cite{sun2017general} \\
144&& 24 && \cite{lin2010new} & & & & & & &  && &&\\
\hline
\end{tabular}}
\end{center}\label{table:large}
\end{table}

\subsection{Projection-based Latin Hypercube Designs}

This subsection reviews several space-filling criteria that emphasize the low-dimensional projection property. The distance-based criteria discussed in Section~\ref{sec:distancelhd} focus solely on space-filling in the full-dimensional space, which can lead to poor projections in lower-dimensional subspaces — an undesirable outcome when only a few factors are active. Constraining these distance-based space-filling designs to the class of Latin hypercube designs can enhance one-dimensional projections but does not guarantee good space-filling properties in higher-dimensional subspaces.  One approach to address this challenge is to use orthogonal array based Latin hypercubes \citep{owen1992orthogonal,tang1993orthogonal}. Alternative, criteria that enforce the projection properties can be used. Below we review orthogonal array-based Latin hypercubes and three  criteria for constructing space-filling designs with projection properties.

Let $S$ be a set of $s$ levels which are taken to be  $1,2,\ldots,s$ in this article. An  $s$-level orthogonal array of $n$ runs, $d$ factors and strength $t$, denoted by $\OA(n,s^d,t)$  is an  $n \times d$ array $\bD$ with entries from $S$ with the property that every $n \times t$ subarray of $\bD$ contains all possible level combinations of $t$ factors exactly $\lambda$ times as a row.     By the definition of orthogonal arrays, a Latin hypercube of $n$ runs for $d$ factors is an $\OA(n,n^d,1)$.

The construction of orthogonal array-based Latin hypercubes, as proposed in  \cite{tang1993orthogonal}, proceeds as follows. Let $\bD$ be an $\OA(n,s^d,t)$. For each column of $\bD$ and $m=1,\ldots,s$, replace the $n/s$ positions containing the entry  $m$ by a random permutation of $(m-1)n/s+1, (m-1)n/s+2,\ldots, mn/s$.
Denote the resulting design after this replacement procedure as $\bL$.   A corresponding orthogonal array-based Latin hypercube design in the space $[0,1)^d$ can then be obtained via (\ref{eq:dij}). In addition to ensuring one-dimensional uniformity,  an $\OA(n,s^d,t)$-based Latin hypercube possesses the $t$-dimensional projection property. Specifically, when projected onto any $t$ columns, the design  places exactly 
   $\lambda=n/s^t$ points in each of the  $s^t$ cells $\mathcal{P}^t$ where $\mathcal{P}=\{[0,1/s],[1/s,2/s), \ldots, [1-1/s,1)\}$.
Example~\ref{exam:oalhs} illustrates the two-dimensional projection property of an $\OA(9,3^4,2)$-based Latin hypercube.   To enhance projection properties, \cite{he2013strong} introduced the concept of strong orthogonal arrays, which serve as a foundation for constructing Latin hypercube designs with superior space-filling characteristics. This has spurred active research on the construction of strong orthogonal arrays; see \cite{chen2024selecting} and references therein for further developments.

\begin{table}[!htb]
\begin{center}
\caption{An $\OA(9,3^4,2)$ and a corresponding orthogonal array-based Latin hypercube}
\begin{tabular}{rrrr rr rrrr}
\multicolumn{5}{c}{ $\bD = \OA(9,3^4,2)$}   & $\quad$ & \multicolumn{4}{c}{ $\bL$}\\
1 &1 &1 &1   &   $\quad$  & $\quad$ & 3 &  3  & 3 &  2    \\
1 &2 &2 &3   &   $\quad$  & $\quad$ &2 &  6  & 6 &  7    \\
1 &3 &3 &2   &   $\quad$  & $\quad$ &1 &  7  & 7 &  6    \\
2 &1 &2 &2   &   $\quad$  & $\quad$ &6 &  2  & 4 &  4    \\
2 &2 &3 &1   &   $\quad$  & $\quad$ &5 &  4  & 9 &  1    \\
2 &3 &1 &3   &   $\quad$  & $\quad$ &4 &  9  & 2 &  9    \\
3 &1 &3 &3   &  $\quad$   & $\quad$ &8 &  1  & 8 &  8    \\
3 &2 &1 &2  &  $\quad$   & $\quad$ & 7 &  5  & 1 &  5    \\
3 &3 &2 &1   &   $\quad$  & $\quad$ & 9 &  8  & 5 &  3    \\
\end{tabular}\label{tab:oa1}
\end{center}
\end{table}

\begin{example}\label{exam:oalhs}
Table~\ref{tab:oa1} presents an orthogonal array-based Latin hypercube, $\bL$, constructed from the orthogonal array $\OA(9,3^4,2)$. Figure~\ref{fig:oalhs1} illustrates the pairwise scatter plot of this Latin hypercube. Each subplot contains exactly one point within each of the nine dot-dash line boxes, demonstrating the design's uniformity in two-dimensional projections.
\begin{figure}[!htb]
\centering
\includegraphics[width=0.85\textwidth]{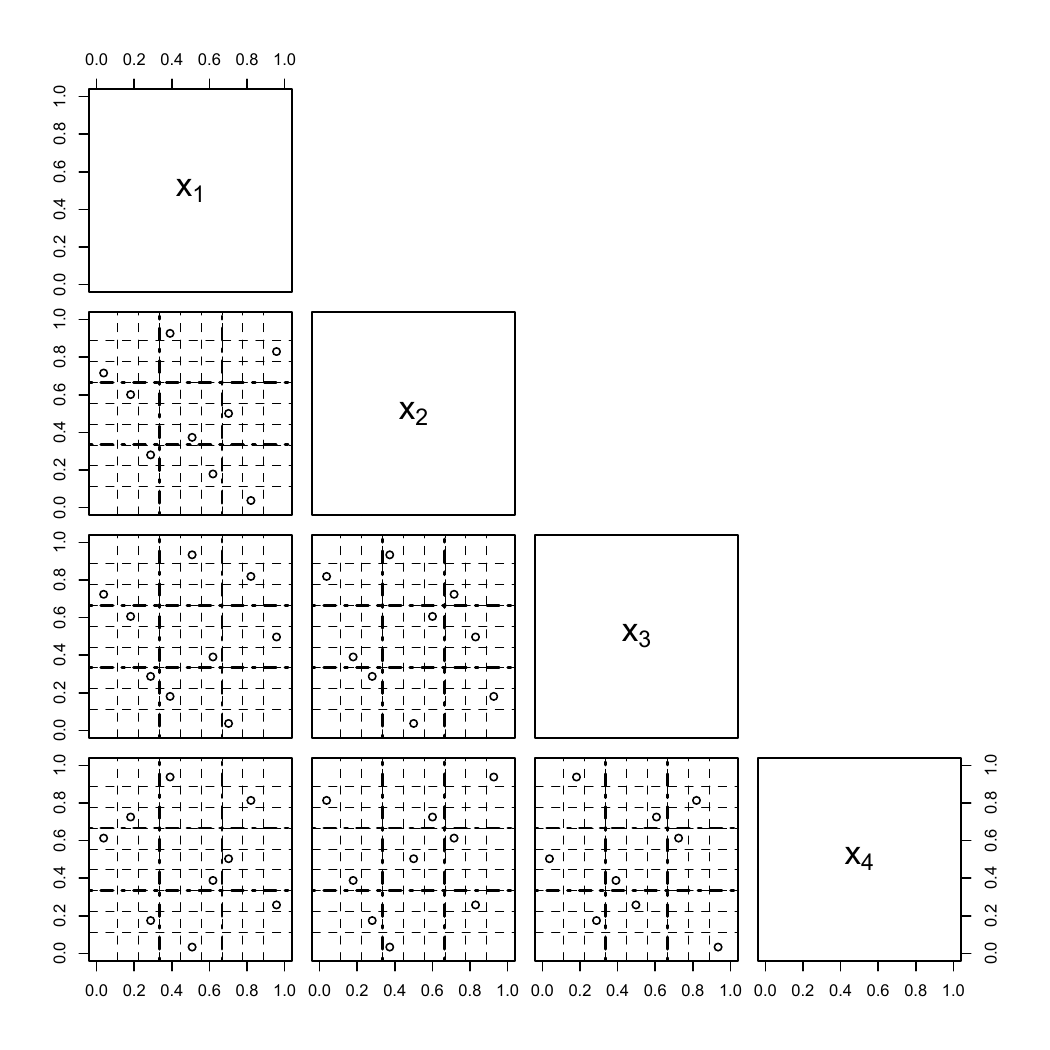}
\caption{The pairwise plot of an $\OA(9,3^4,2)$-based Latin hypercube design  for the four factors $X_1,\ldots, X_4$ }\label{fig:oalhs1}
\end{figure}
\end{example}

Now we review three projection-based criteria for constructing space-filling designs.

{\bf Minimum Average Reciprocal Distance Criterion.} 
\cite{draguljic2012noncollapsing} introduced the minimum average reciprocal distance (ARD) criterion which aims to minimize
\begin{equation} \label{eq:ard}
ARD(\bm X)  =  \left \{ \frac{1}{
\binom{n}{2} \sum_{k \in J }
\binom{d}{k}}
\sum_{k \in J} \sum_{|\mathcal{U}| =k} \sum_{i=1}^{n-1} \sum_{j=i+1}^n \left( \frac{k^{1/2}}{d_2^{1/2}(\bm x_{i,\mathcal{U}}, \bm x_{j,\mathcal{U}})} \right )^\lambda
\right \}^{1/\lambda}
\end{equation}
\noindent for an $n \times d$ design matrix ${\bm X} = (x_{ij})$, where $\lambda \geq 1$ is a prespecified real number, $J$ and $\mathcal{U}$ are subsets of $\{1,2,\ldots,d\}$ and $|\mathcal{U}|$ represents the cardinality of $\mathcal{U}$, $\bm x_{i,\mathcal{U}}$ and $\bm x_{j,\mathcal{U}}$ are the $i$th and $j$th runs of ${\bm X}$ projected onto dimensions indexed by the elements of $\mathcal{U}$.

{\bf Maximum Projection Criterion.}  \cite{joseph2015} proposed the maximum projection (MaxPro) Latin hypercube designs that consider designs’ space-filling properties in all possible dimensional spaces. Such designs minimize the maximum projection criterion, which is defined as:
\begin{equation}\label{eq:maxpro}
     \psi(\bm X) = \left( \frac{1}{\binom{n}{2}} \sum_{i=1}^{n-1} \sum_{j=i+1}^{n} \frac{1}{\prod_{l=1}^{d} (x_{il} - x_{jl})^2} \right)^{1/d}.
\end{equation}
One can see that any two design points should be apart from each other in any projection to minimize the value of $\psi(\bm X)$.

{\bf Uniform Projection Criterion.}  In the same spirit as \cite{joseph2015}, \cite{sun2019uniform} 
introduced the uniform projection (UP) criterion defined as 
$$ \phi(\bm X) = \frac{2}{d(d-1)} \sum_{|\mathcal{U}|=2} CD(\bm X_{\mathcal{U}})$$
\noindent where 
$CD(\bm X_{\mathcal{U}})$ is the squared centered $L_2$ discrepancy of $\bm X_{\mathcal{U}}$, the projected design of $\bm X$ onto dimensions indexed by the elements of $\mathcal{U}$, $\mathcal{U} \in \{1,2,\ldots, d\}$ and $|\mathcal{U}|$ stands for  the cardinality of $\mathcal{U}$. The squared centered $L_2$ discrepancy of an $n \times d$ design $\bm X =(x_{ij})$ each column of which has $s$ levels is defined as
$$CD(\bm X)  = \frac{1}{n^2} 
\sum_{i=1}^n \sum_{j=1}^n \prod_{k=1}^d (1 + \frac{1}{2}|z_{ik}| + \frac{1}{2}|z_{jk}| - \frac{1}{2}|z_{ik}-z_{jk}|)-
\frac{2}{n} \sum_{i=1}^n\prod_{k=1}^d ( 1 + \frac{1}{2}|z_{ik}| -
 \frac{1}{2}|z_{ik}|^2) + (\frac{13}{12})^d, 
$$
\noindent with $z_{ik} = (2x_{ik} - s +1)/(2s)$.

\begin{example}\label{example:projection}
We examine the projection properties of maximin, ARD, MaxPro, and UP Latin hypercube designs using a design of 10 runs with four input variables.
The maximin Latin hypercube design is selected as the best from 100 maximin 
$L_2$-distance Latin hypercubes generated using the \pkg{SLHD} package in \proglang{R} \citep{ba2015optimal} with default settings. The ARD Latin hypercube design is constructed with $\lambda =1$ and $J = \{1,2\}$. The MaxPro Latin hypercube design is obtained by running the {\em MaxProLHD} function from  the \pkg{MaxPro} package in \proglang{R} \citep{joseph2015} 100 times with default settings. Similarly, the UP Latin hypercube design is generated using a threshold accepting algorithm \citep{sun2019uniform,wang2022design}.
Figure~\ref{fig:projection} presents the pairwise scatter plots of the four input variables for these designs. Visually, the RD, MaxPro, and UP Latin hypercube designs exhibit slightly better two-dimensional projection properties than the maximin Latin hypercube design. Notably, in the maximin design, the scatter plot of 
$X_1$ and $X_4$ reveals points aligned along two lines, a pattern absent in the other three designs.
Computing the criterion (\ref{eq:ard}) with 
$J = 2$ and $\lambda = 1$ yields values of 0.306, 0.292, 0.297, and 0.292 for the maximin, ARD, MaxPro, and UP Latin hypercube designs, respectively.

\begin{figure}[!htb]
     \centering
     \begin{subfigure}[b]{0.45\textwidth}
         \centering
         \includegraphics[width=\textwidth]{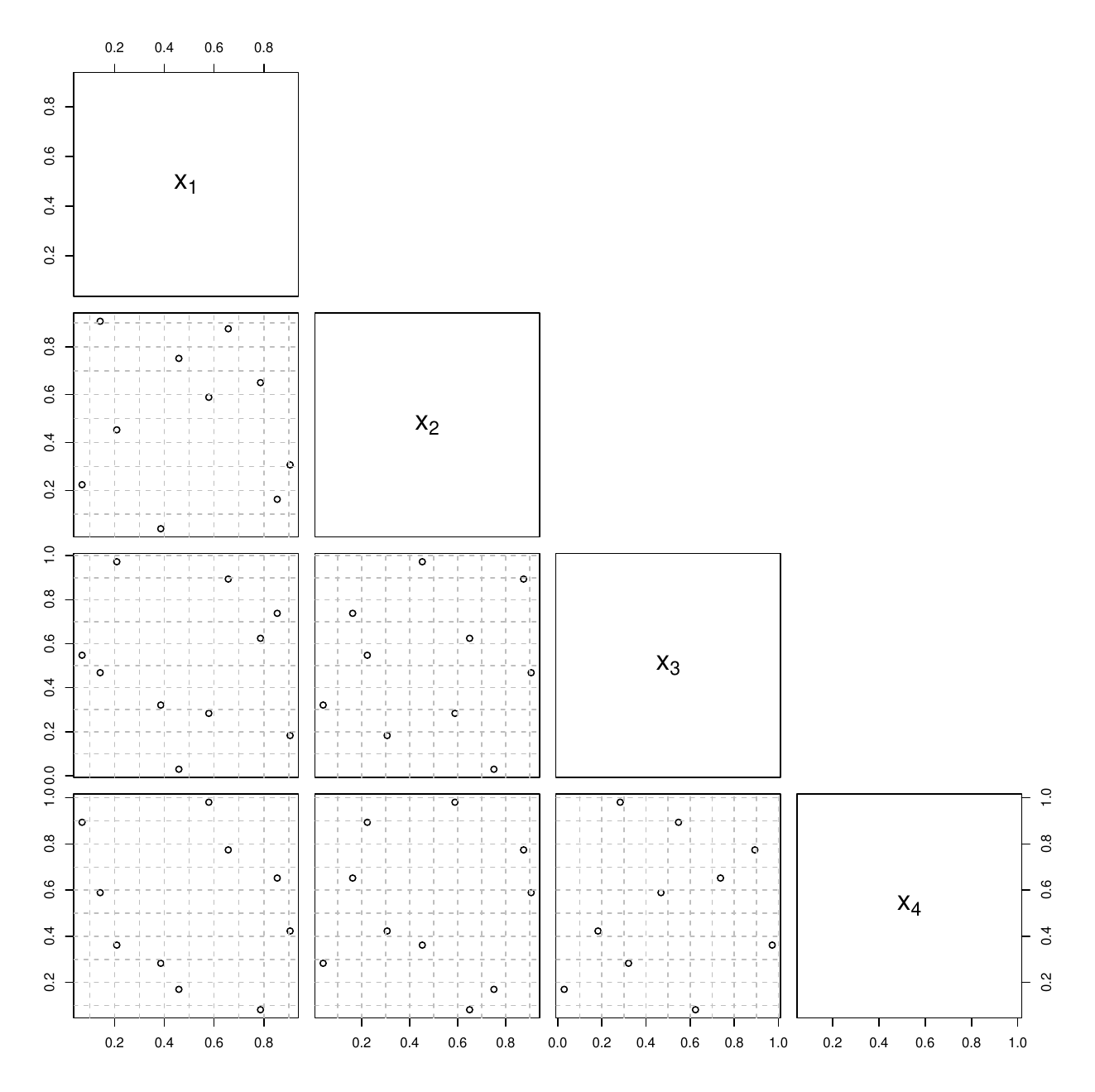}
         \caption{Maximin  }
         \label{fig:1}
     \end{subfigure}
     \hfill
     \begin{subfigure}[b]{0.45\textwidth}
         \centering
         \includegraphics[width=\textwidth]{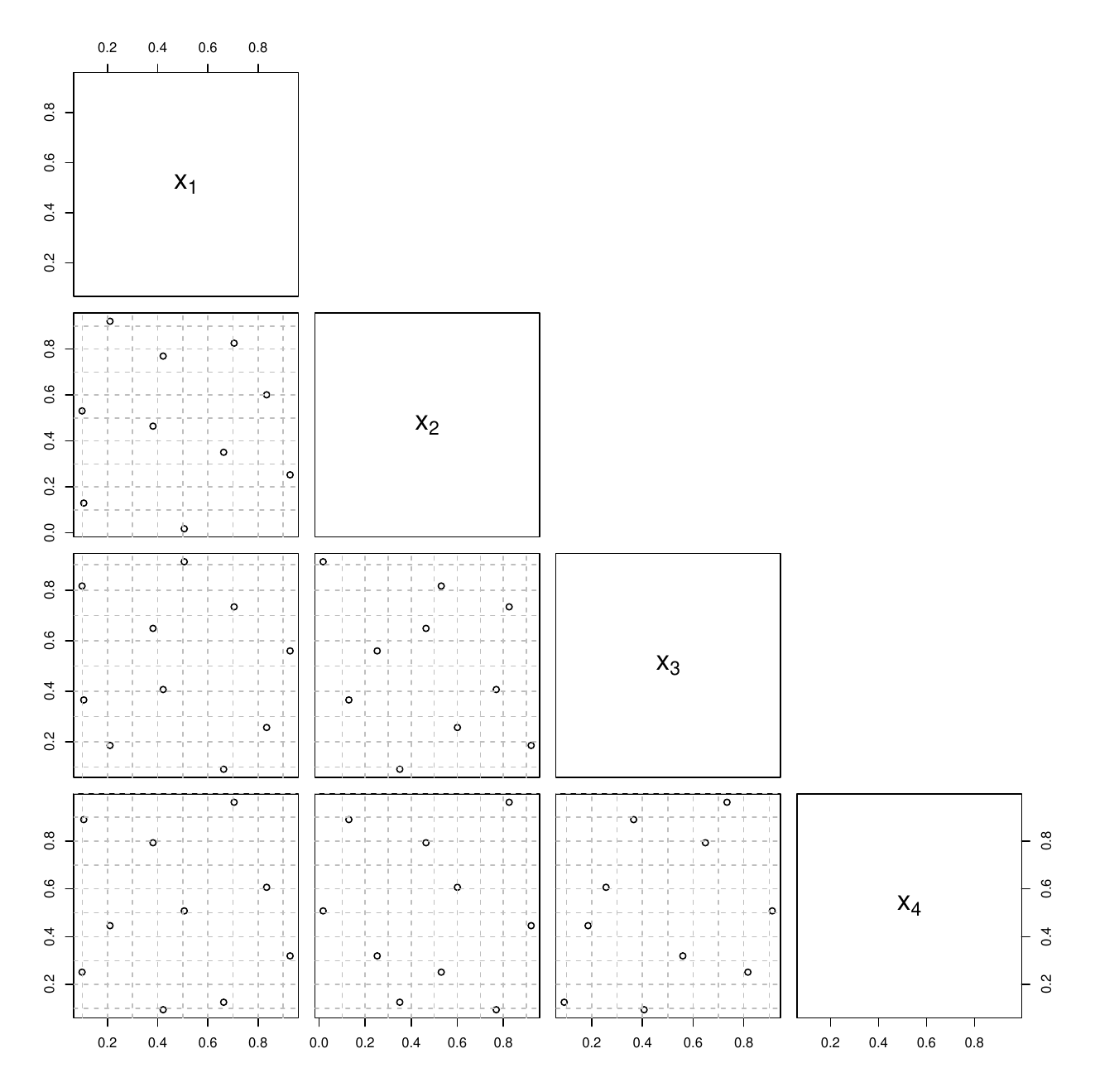}
         \caption{ARD  }
         \label{fig:2}
     \end{subfigure}
     \begin{subfigure}[b]{0.45\textwidth}
         \centering
         \includegraphics[width=\textwidth]{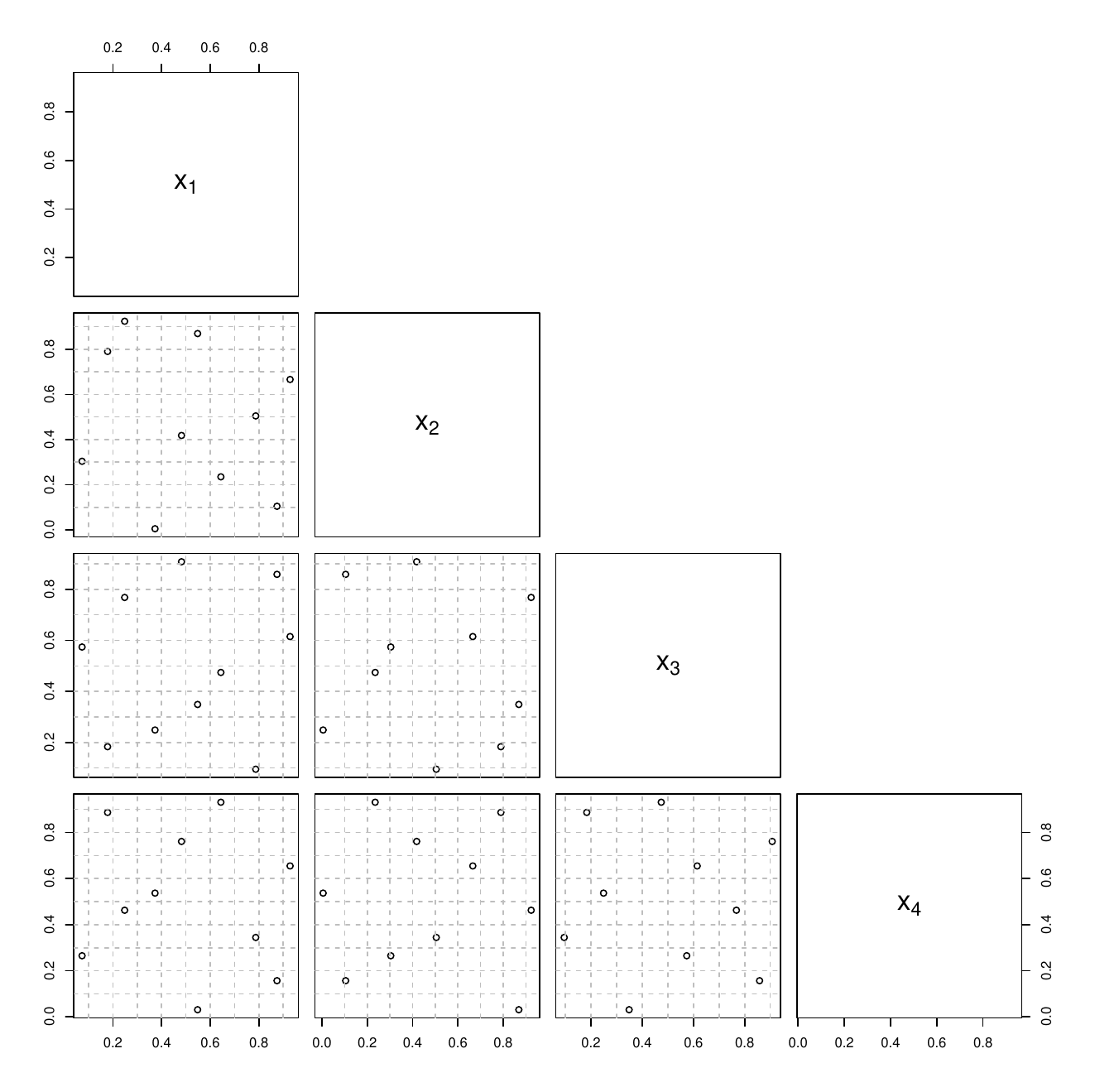}
         \caption{MaxPro }
         \label{fig:3}
     \end{subfigure}
          \hfill
     \begin{subfigure}[b]{0.45\textwidth}
         \centering
         \includegraphics[width=\textwidth]{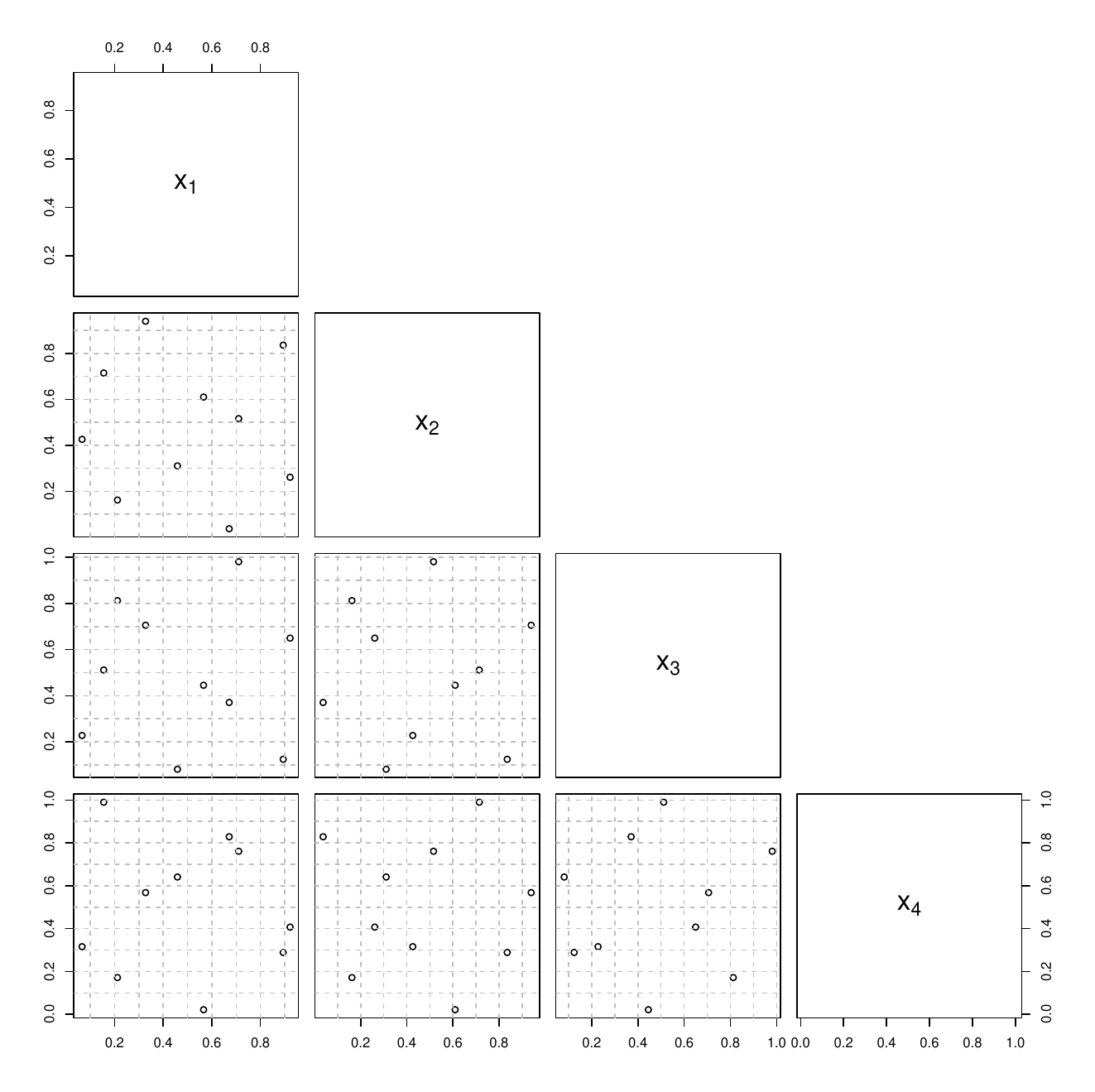}
         \caption{UP  }
         \label{fig:4}
     \end{subfigure}
        \caption{The pairwise plot of the Maximin, ARD, MaxPro, UP Latin hypercube designs of 10 runs for the four input variables  $X_1,X_2,X_3,X_4$ in Example~\ref{example:projection}}
        \label{fig:projection}
\end{figure}
\end{example}

\subsection{Space-filling Designs Beyond Latin Hypercubes}


This subsection explores various space-filling designs beyond Latin hypercube designs. These include space-filling designs with many levels, low-discrepancy sequences, space-filling designs for constrained design spaces, and
space-filling designs tailored for different types of computer experiments.

\subsubsection{Designs with Many Levels}
\cite{bingham2009orthogonal} argued that in many practical applications, it is not necessary for the number of levels to equal the run size. Instead, they proposed the use of orthogonal and nearly orthogonal designs for computer experiments, which allow factors to have repeated levels while maintaining zero or low correlations. This approach encompasses a broad class of orthogonal designs, including two-level orthogonal designs and orthogonal Latin hypercubes as special cases. For construction methods of orthogonal and nearly orthogonal designs, see \cite{bingham2009orthogonal} and \cite{georgiou2014construction}. Table~\ref{tab:od} presents a $4 \times 2$ orthogonal design $\bD_1$ and a $8 \times 4$ orthogonal design $\bD_2$ where in $\bD_1$, $x_1$ and $x_2$ is a permutation of $\{1,2\}$, and in $\bD_2$, $x_1,x_2,x_3,x_4$ is a permutation of $\{1,2,3,4\}$.

\begin{table}[!htb]
\begin{center}
\caption{A $4 \times 2 $ orthogonal design and a $8 \times 4$ orthogonal design }
\begin{tabular}{rrrrrrr}
\multicolumn{2}{c}{$n=4$} & &  \multicolumn{4}{c}{$n=8$}\\ 
$x_1$ & $x_2$  &   $\quad$ & $x_1$ & $-x_2$ & $x_4$ & $x_3$\\
$-x_1$ & $-x_2$  &   $\quad$ & $x_2$ & $x_1$ & $x_3$ & $-x_4$\\
$x_2$ & $-x_1$  &   $\quad$ & $x_3$ & $-x_4$ & $-x_2$ & $-x_1$\\
$-x_2$ & $x_1$  &   $\quad$ & $x_4$ & $x_3$ & $-x_1$ & $x_2$\\
&& $\quad$ & $-x_4$ & $-x_3$ & $x_1$ & $-x_2$\\
&& $\quad$ &$-x_3$ & $x_4$ & $x_2$ & $x_1$\\
&& $\quad$ & $-x_2$ & $-x_1$ & $-x_3$ & $x_4$\\
&& $\quad$ & $-x_1$ & $x_2$ & $-x_4$ & $-x_3$\\
\end{tabular}\label{tab:od}
\end{center}
\end{table}

\subsubsection{Low-discrepancy Sequence}

\begin{figure}[!htb]
\centering
\includegraphics[width=0.85\textwidth]{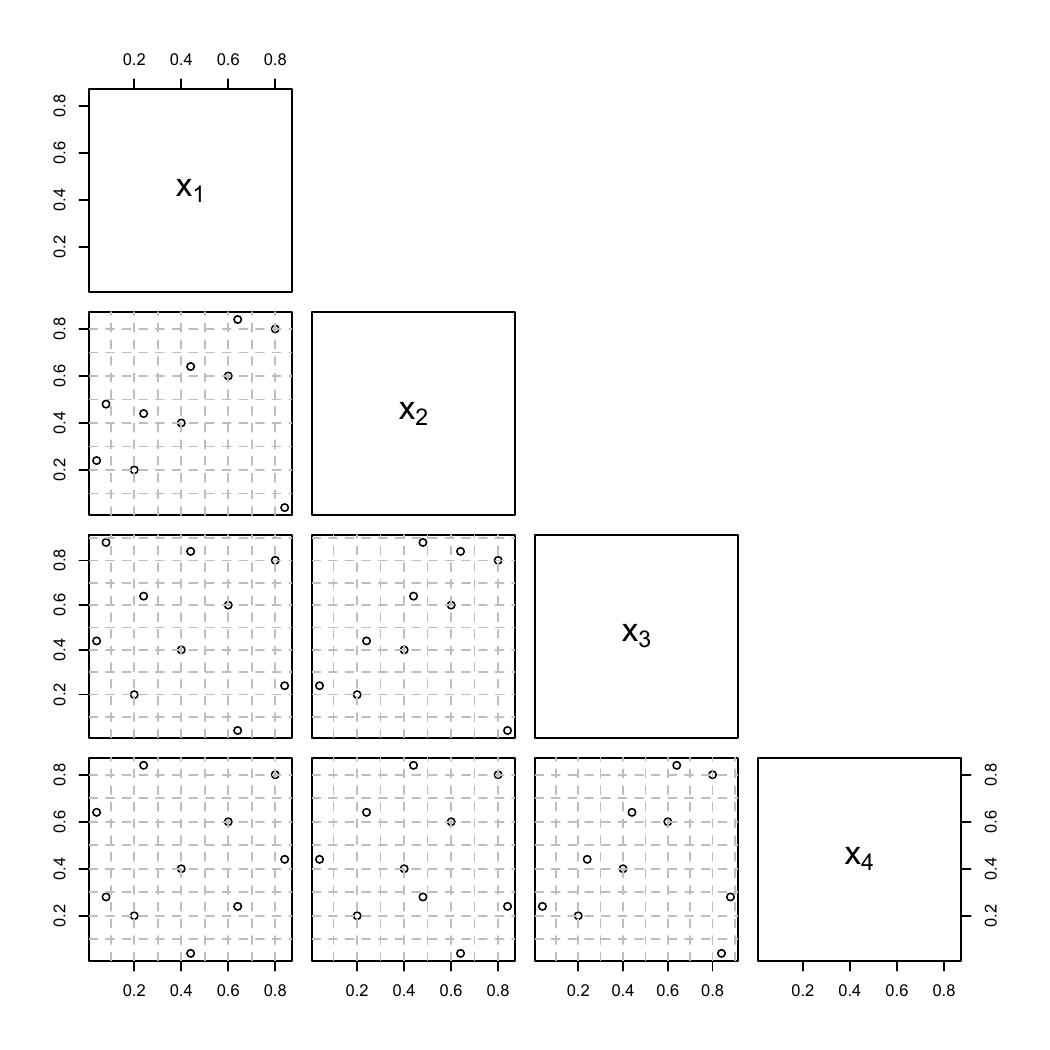}
\caption{The pairwise plot of a low discrepancy Faure sequence  of 10 points for the four factors $X_1, X_2, X_3, X_4$ }\label{fig:faure}
\end{figure}

Another widely used class of space-filling designs is quasi-random low-discrepancy sequences, which measure uniformity based on discrepancy, or the deviation from a perfectly uniform distribution. These sequences originate from numerical integration problems, where the goal is to estimate the expectation $\mu$ of 
 $y = f(\bx)$ with  $f$ being known and $\bx =(x_1, \ldots, x_d)$ having a uniform distribution in the design space $\mathcal{X} = [0,1)^d$ and $y \in R$.  Given a set of $n$ points in the design space $\mathcal{X}$, and using $\hat{\mu}$ to approximate $\mu$, the integration error bound is determined by the Koksma-Hlawka inequality,
$$| \mu - \hat{\mu}| \leq V(f) D^*(\mathcal{X}),$$
\noindent where  $V(f)$ is the variation of $f$ in the sense of Hardy and Krause and $D^*(\mathcal{X})$ is the star discrepancy,
$$D^*(\mathcal{X}) = \max_{\bx \in \mathcal{X}} \left |
 \frac{N(\mathcal{X},J_{\mathbf x})}{n} -\hbox{Vol}({J_{\mathbf x}}) 
\right |, $$
\noindent with $ J_{\mathbf x} =[0,{\mathbf x})$ being the interval of $[0,x_1)\times
[0,x_2) \times \cdots \times [0,x_d)$, $N(\mathcal{X},J_{\mathbf x})$ being the number of points of $\mathcal{X}$ falling in $J_{\mathbf x}$, and $\hbox{Vol}(J_{\mathbf x})$ being the volume if the interval $J_{\mathbf x}$.  The star discrepancy measures how well the points in $\mathcal{X} $ approximate the uniform distribution by comparing the proportion of points in subregions of  $\mathcal{X}$ to their theoretical volumes. 
Quasi-Monte Carlo methods seek to generate points that minimize this discrepancy. A sequence $S$ is considered a {\em low-discrepancy sequence} if its first 
$n$ points satisfy
$$D^*(\mathcal{X}) = O(n^{-1}\hbox{log} (n)^d),$$
\noindent where $O(\cdot)$ is the big O notation. That is, a low-discrepancy sequence ensures better uniformity compared to purely random sampling. Well-known examples include Halton, Hammersley, Sobol sequences, ($t,s)$-sequences, ($t,m,s$)-nets and uniform designs, along with their numerous variants \citep{halton1960efficiency,sobol1967distribution}.  QMCPy is a Python library designed for Quasi-Monte Carlo methods \citep{QMCPy2020a}. The library provides a range of functionalities for generating low-discrepancy sequences. 
Figure~\ref{fig:faure} presents the pairwise plot of a 10-point low-discrepancy Faure sequence (generated using the {\em runif.faure} function in \pkg{DiceDesign} package in \proglang{R}  \citep{DiceDesign}) across four input variables. The plot highlights a key limitation of such sequences — while they achieve uniform coverage of the design space, their low-dimensional projections may not exhibit desirable space-filling properties. This suggests that despite their theoretical advantages in discrepancy minimization, low-discrepancy sequences may still suffer from uneven distribution in lower-dimensional subspaces.
This motivates the uniformity measures such as symmetric $L_2$ discrepancy, centered $L_2$ discrepancy, and modified $L_2$ discrepancy, to enforce the projection
uniformity in low dimensions \citep{hickernell1998generalized}. 

The construction of low-discrepancy sequences remains an active area of research within quasi-Monte Carlo methods, with several approaches available, such as the good lattice point method, good point method, Halton sequences, Faure sequences, $(t,s )$-sequences and uniform designs. For a comprehensive overview of low-discrepancy sequences and their applications, see \cite{niederreiter1992random}.

\subsection{Space-filling Designs for Constrained Design Space}

Many scientific, engineering, and applied fields often involve nonrectangular constrained input regions, where input variables are subject to linear or nonlinear constraints or exhibit probabilistic relationships and dependencies  \citep{iman1982distribution,stinstra2003constrained,petelet2010latin,bowman2013weighted,kang2019stochastic}. 
These constraints can arise naturally from physical, geometric, or operational limitations. For example, \cite{draguljic2012noncollapsing} illustrated this challenge with a biomedical engineering application, where a computer model is used to evaluate the mechanical performance of prosthetic devices. The model incorporates four input variables: tip displacement ($X_1$), rotation of the implant axis about the lateral axis at the tip ($X_2$), rotation of the implant axis about the anterior axis at the tip ($X_3$), and rotation about the implant axis ($X_4$). These variables are constrained by the following specific relationships,
$$ -10 \leq 5X_2+ 2X_3 \leq 10,$$
$$ -10 \leq -5X_2 + 2X_3 \leq 10.$$
\noindent That is, the input variables $X_2$ and $X_3$ is restricted to a diamond-shaped region. 
Addressing such constraints in experimental designs is crucial for generating feasible and representative simulations.   Simply removing infeasible points from a design can lead to an insufficient sample size and non-uniform point distribution, making it challenging to achieve desired space-filling properties. To address this, constrained experimental designs have been developed. \cite{draguljic2012noncollapsing} proposed an algorithm for constructing space-filling designs in high-dimensional linear constrained regions, ensuring a well-distributed sample while satisfying constraints. For bounded convex regions, iterative clustering-based methods have shown promising results, including the hierarchical clustering approach by \cite{lekivetz2015fast}. To handle irregular experimental regions, \cite{chen2019optimal} introduced a discrete particle swarm optimization method to generate optimal space-filling designs. \cite{kang2019stochastic} proposed a stochastic coordinate-exchange algorithm to construct space-filling designs (and optimal designs for physical experiments) for any irregular shaped design space, including both convex and non-convex, or even disconnected constrained domains by projecting the original $d$-dimensional constraints into 1-dimensional space due to coordinate-exchange algorithm. 

Beyond geometric constraints, some applications require incorporating probabilistic dependencies among input variables. \cite{bowman2013weighted} developed weighted space-filling designs, allowing users to specify a weight function to reflect known multivariate dependencies in simulators. \cite{joseph2015sequential} introduced the minimum energy design, where design points behave as charged particles, and their total potential energy is minimized to follow a target probability density function. Expanding on these ideas, \cite{lu2021non} proposed non-uniform space-filling designs, which integrate weighted space-filling and minimum energy designs to allow flexible, non-uniform density distributions throughout the input space. These advancements enable more effective experimental designs for complex and constrained simulation studies.

\subsection{Space-filling Designs for Advanced Computer Experiments}


Space-filling designs discussed so far are well-suited for basic computer experiments with only quantitative inputs. However, more complex scenarios — such as multi-fidelity experiments, multiple computer experiments, or those involving both quantitative and qualitative inputs — required specialized design strategies.
In many applications, computer codes can be executed at varying levels of accuracy, leading to multi-fidelity computer experiments \citep{goh2013prediction}. To accommodate this, \cite{qian2009nested} proposed {\em nested space-filling designs}, where the full set of design points is used for low-fidelity experiments, while a carefully selected subset is allocated for high-fidelity experiments. Crucially, both the full set and the subset maintain space-filling properties, ensuring efficient exploration across fidelity levels.

In addition, investigators often work with multiple computer codes for the same system due to the diversity of mathematical models and numerical methods available \citep{yang2013construction}. To facilitate multiple computer experiments, \cite{qian2009sliced} introduced sliced space-filling designs, where each slice is used for a separate experiment. A sliced space-filling design retains its space-filling property while being divisible into multiple smaller space-filling designs. The construction of sliced Latin hypercube designs remains an active area of research.

In practical computer experiments, both quantitative and qualitative input variables are often involved \citep{qian2008gaussian, deng2017additive}. Sliced space-filling designs can be utilized in such cases, where each slice corresponds to a level combination of qualitative input variables. However, as the number of qualitative variables increases, this approach becomes inefficient due to the large run size required.
To achieve greater run-size efficiency and flexibility, \cite{deng2015design} introduced marginally coupled designs, which integrate an orthogonal array for qualitative inputs with a space-filling design for quantitative inputs. Specifically, the design for quantitative inputs maintains space-filling properties within each level of the qualitative variables, effectively functioning as a sliced space-filling design with respect to each qualitative input variable. This approach allows for a more economical run size while preserving desirable design properties. Further research has been devoted to the construction of marginally coupled designs and their refinements, as seen in \cite{he2017construction} and \cite{yang2023doubly}.

Beyond accommodating different data structures in computer experiments, a recent development in space-filling designs focuses on incorporating prior knowledge. Standard space-filling designs are ideal when there is no prior knowledge on how input variables influence responses, particularly in cases where the primary objective is global fitting. However, when prior information about the system’s true process or the suitability of specific statistical models is available, a natural question arises: Can more effective designs be constructed than standard space-filling designs?
\cite{chen2025construction} initiated research in this area by considering scenarios where factors are grouped into disjoint sets with no interactions between them. In such cases, the underlying process or preferred surrogate model follows an additive structure, where each component function depends only on a specific group of variables. To leverage this structure, \cite{chen2025construction} proposed {\em grouped orthogonal arrays}-based space-filling designs, which improve low-dimensional projections within each group. Empirical studies demonstrated that grouped orthogonal arrays-based designs enhance predictive accuracy compared to traditional space-filling approaches. Several construction methods were developed, and many designs were tabulated to facilitate practical implementation.

  

\section{Numerical Illustrations}

This section presents an empirical comparison of various design approaches, including maximin Latin hypercube designs, orthogonal array-based Latin hypercube designs, ARD, UP and MaxPro Latin hypercube designs, and distance-distributed designs \citep{zhang2021distance}.  The latter was proposed motivated by the observation that maximin Latin hypercube designs perform worse than random designs in estimating the correlation parameters. They study the distribution of pairwise distances between design points, and developed
a numerical scheme to optimize those distances for a given sample size and dimension. They name the proposed design {\em distance-distributed} designs and use `lhsbeta` to label their designs. 
For clarity, the designs under comparison are labeled as `maximinLHD', `OALHS', `ARD', `UP', `MaxPro', and `lhsbeta', respectively. All implementations in this section utilize the \pkg{DiceKriging} package \citep{DiceKriging} in \proglang{R} for model fitting and prediction. The chosen kernel function for kriging is Matern(3/2). To compare the prediction accuracy, we generate a test set $\mathcal{W}$ of $N$ points and compute the RMSPE defined as
\begin{equation*} 
RMSPE = \sqrt{\frac{1}{N}\sum_{\{ \bx_i \in \mathcal{W},i=1,\ldots,N\}}(\hat{y}(\bx_i) - y(\bx_i))^2  }
\end{equation*}
Here, we consider random Latin hypercube designs of $N=5,000$ points. We begin with a well-known computer simulator, the Robot Arm function, which is widely used in neural network research. Example \ref{ex:robot} provides an empirical comparison of six design approaches, evaluating their performance in terms of predictive accuracy.

\begin{example}\label{ex:robot}
The Robot Arm simulator models the position of a robot arm which has four segments. While the shoulder is fixed at the origin, the four segments each have length $L_i$, and are positioned at an angle $\theta_i$ for $i=1,2,3,4$.  The computer simulator is represented by
$$f(\bx) = (u^2+v^2)^{0.5},$$
$$ u = \sum_{i=1}^4 L_i cos(\sum_{j=1}^i \theta_j),$$
$$ v = \sum_{i=1}^4 L_i sin(\sum_{j=1}^i \theta_j). $$
The run sizes considered in this study are 81, 162, and 256, selected to ensure the availability of orthogonal arrays for constructing orthogonal array-based Latin hypercube designs. Specifically, we use orthogonal arrays with 81 runs and 9 levels, 162 runs and 9 levels, and 256 runs and 16 levels. These orthogonal arrays are taken from the website \url{http://neilsloane.com/oadir/}.
Figure~\ref{fig:robot} presents the mean and standard deviation of log RMSPEs for the `maximinLHD', `OALHS', `ARD', `UP', `MaxPro', and `lhsbeta' methods, evaluated over 50 simulations using the Robot Arm computer simulator. The results indicate that for smaller designs, 'lhsbeta' achieves the lowest RMSPE, while `maximinLHD' performs the worst, consistent with findings from \cite{zhang2021distance}. For larger run sizes, the `UP' approach yields the most accurate predictions overall, whereas `OALHS' demonstrates the poorest performance.
\begin{figure}[!htp]
    \centering
      \includegraphics[width=8.1cm]{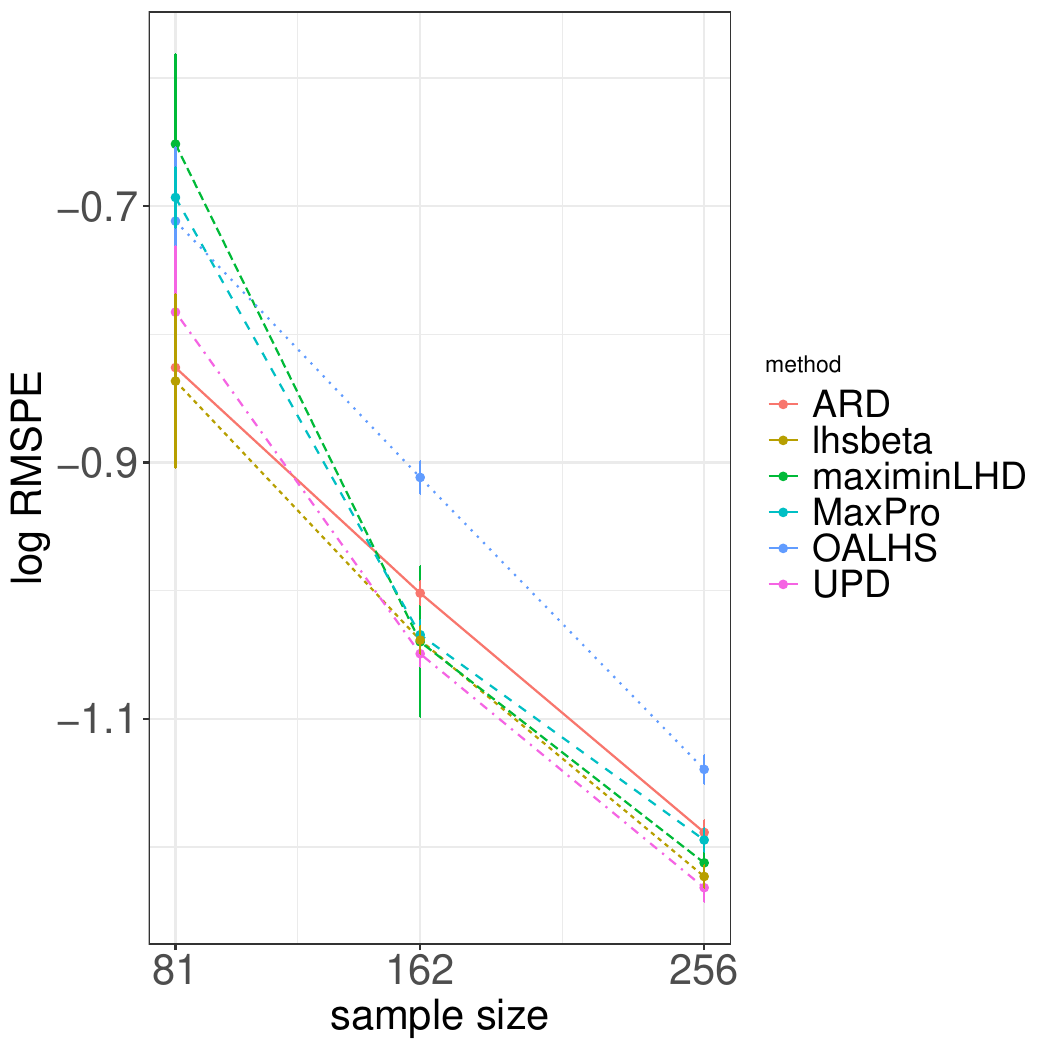}
   \caption {Mean and standard deviation of log RMSPEs associated with the `maximinLHD', `OALHS', `ARD', `UP', `MaxPro', `lhsbeta' methods over 50 simulations for the Robot arm computer simulator }\label{fig:robot}
\end{figure}

\end{example}

The run sizes for orthogonal array-based Latin hypercube designs are limited. Thus, in Example \ref{ex:gf}, we exclude  the  `OALHS' method.  Example \ref{ex:gf} provides the empirical comparison among  the `maximinLHD',  `ARD', `UP', `MaxPro', `lhsbeta' methods in the four computer models with 3, 5, 6, 9 input variables respectively. The sample sizes considered are 50,100,200, and 300. 
\begin{example}\label{ex:gf} 
We consider the following four functions:
\begin{itemize}
    \item[(a)] the detpep10 function \citep{dette2010generalized}, $d=3$;
        \item[(b)] the Friedman \citep{friedman1991multivariate} function, $d=5$; 
    \item[(c)]  the GramacyLee function \citep{gramacy2009adaptive}, $d=6$;
    \item[(d)] the Bratley function \citep{bratley1992implementation}, $d=9$. 
\end{itemize}

\begin{figure}[!htp]
  \begin{subfigure}{0.47\textwidth}
      \centering
      \includegraphics[width=\textwidth]{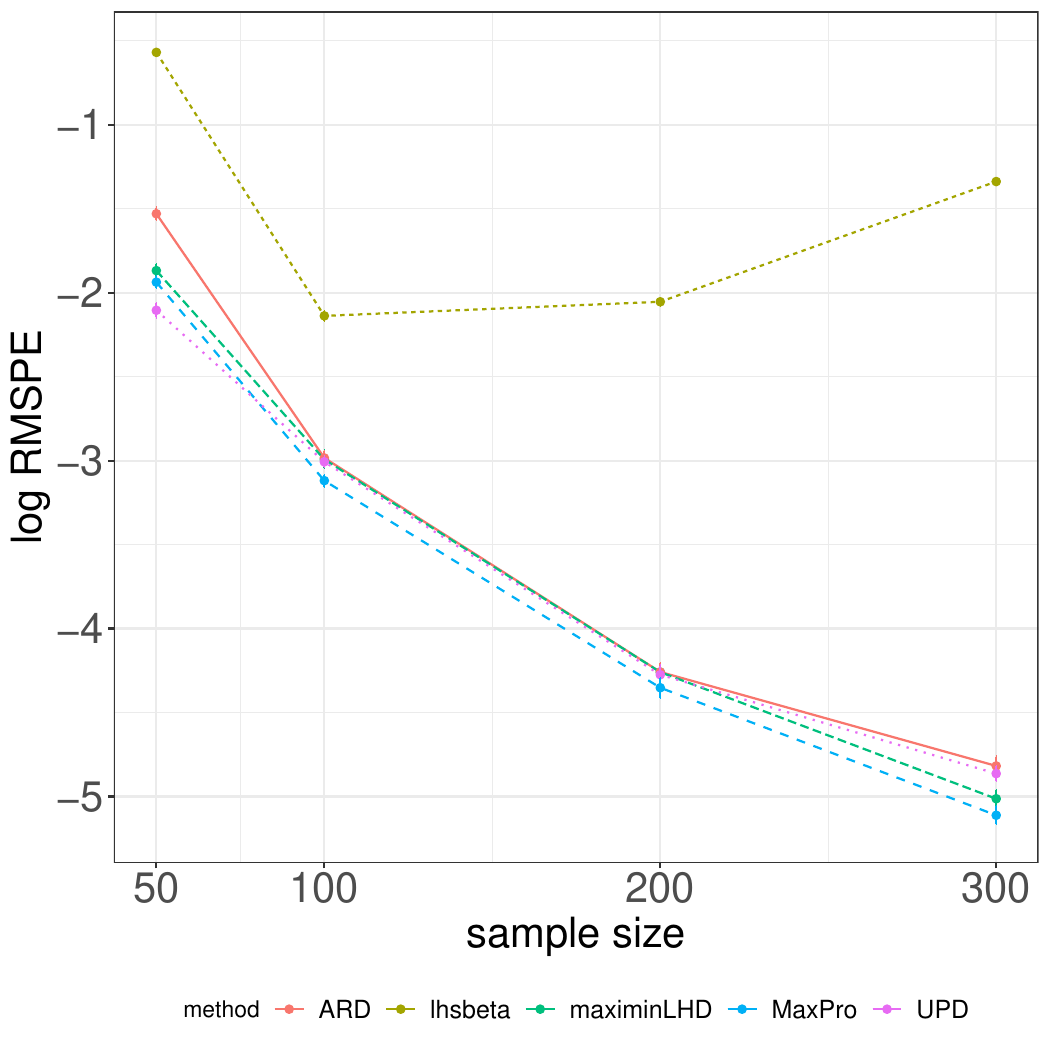}
      \caption{the detpep10 function, $d=3$}\label{fig:2a}
  \end{subfigure}%
   \begin{subfigure}{0.47\textwidth}
      \centering
      \includegraphics[width=\textwidth]{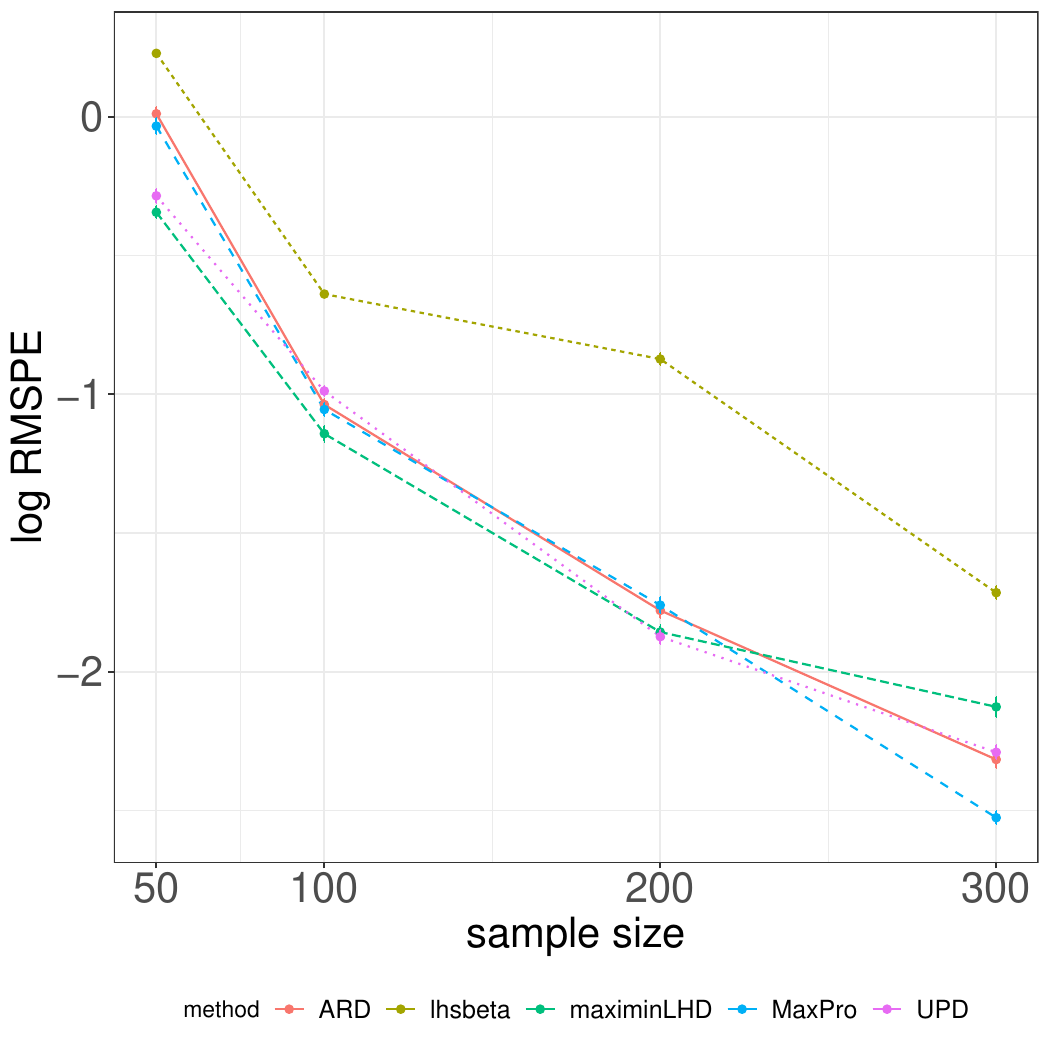}
      \caption{the Friedman function, $d=5$}\label{fig:2b}
 \end{subfigure}%
 \\
  \begin{subfigure}{0.47\textwidth}
      \centering
      \includegraphics[width=\textwidth]{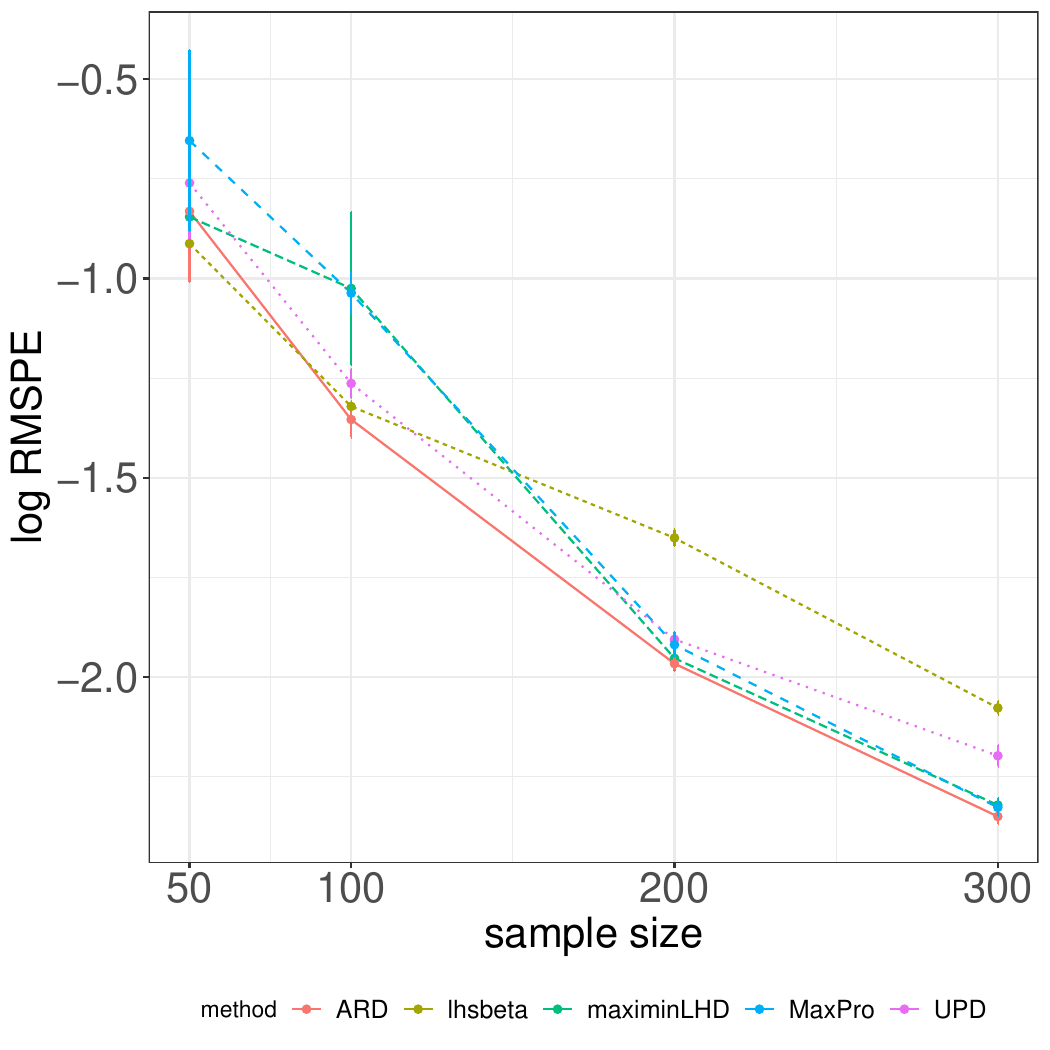}
      \caption{the GramacyLee function, $d=6$}\label{fig:2c}
    \end{subfigure}
   \begin{subfigure}{0.47\textwidth}
      \centering
      \includegraphics[width=\textwidth]{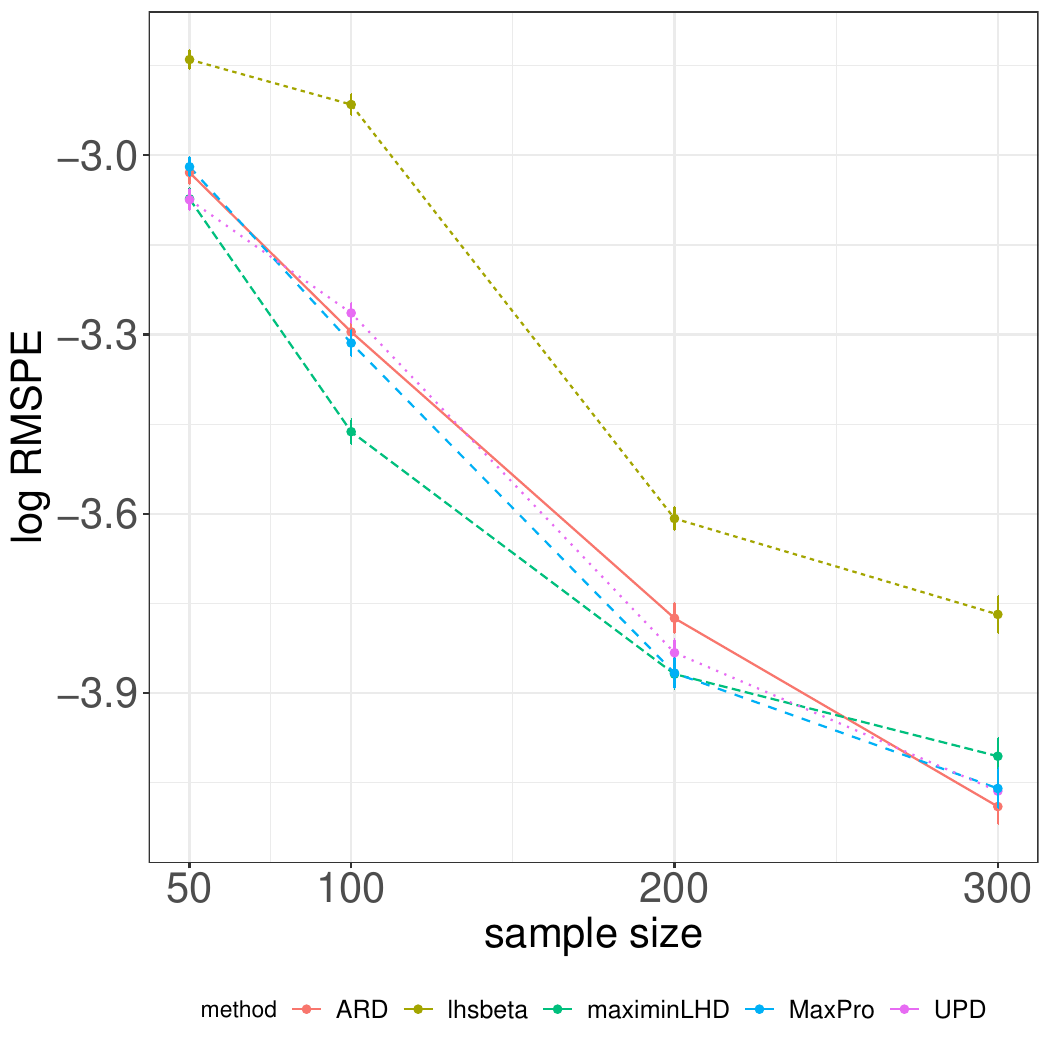}
      \caption{the Welch function, $d=9$}\label{fig:2d}
\end{subfigure}
  \caption {Mean and standard deviation of log RMSPEs associated with the `maximinLHD', `ARD', `UP', `MaxPro',`lhsbeta' methods over 50 simulations for the   detpep10, Friedman, GramacyLee, and Bratley  functions  }\label{fig:2comp}
\end{figure}

Figure~\ref{fig:2comp} presents the mean and standard deviation of log RMSPEs for the `maximinLHD', `ARD', `UP', `MaxPro', and `lhsbeta' design methods, evaluated across 50 simulations using four computer simulators. The results suggest that no single design consistently outperforms the others across all simulators and sample sizes. Among the methods considered, the `lhsbeta' designs exhibit the highest prediction errors. However, in the context of log RMSPE scale, the `lhsbeta' method still achieves reasonably accurate predictions.

\end{example}

Through these empirical investigations, we find that optimal space-filling designs, evaluated using different criteria, yield comparable performance in global fitting with Gaussian process emulators. No single class of space-filling designs is universally optimal. From a computational standpoint, generating `ARD' and `UP' Latin hypercube designs is significantly more efficient than producing `MaxPro', `maximinLHD', and `lhsbeta' designs. Table~\ref{tab:time} reports the computation time required to obtain the ``optimal'' design from 100 `maximinLHD', `ARD', `UP', MaxPro', and `lhsbeta' designs, respectively, on an Apple laptop with M2 chip. Notably, the computational cost of lhsbeta' remains stable as the number, $d$, of input variables
 increases. In contrast, for other methods, computation time grows substantially with increasing $d$. 


\begin{table}[!htb]
\begin{center}
\caption{Computation time (in seconds) of the `maximinLHD', `ARD', `UP', `MaxPro', and `lhsbeta' designs}
\begin{tabular}{rrrrrrr}
$n$ &$\quad $ $d$ $\quad $ & maximinLHD &  ARD& UP& MaxPro&lhsbeta\\
\hline
\multirow{3}{1em}{ 50}
& $\quad $ 3 $\quad$ &39.846  & 2.216&4.212 & 16.227& 29.726\\
& $\quad $ 6 $\quad $ & 51.361 &6.197 &5.079 & 27.222& 30.310\\
& $\quad $ 9 $\quad $ & 59.607 & 12.768&5.683 & 35.484& 30.856\\
& &&&&&\\
\multirow{3}{1em}{100}
& $\quad $ 3 $\quad $ &174.245  & 4.928&11.003 & 66.532& 109.798\\
& $\quad $ 6 $\quad $ & 242.75 &15.262 &14.281 &136.112&  108.105\\
& $\quad $ 9 $\quad $ & 299.233 & 31.897 &16.998 & 190.634& 112.259\\
& &&&&&\\
\multirow{3}{1em}{200}
& $\quad $ 3 $\quad $ &754.518  & 13.591& 33.752 &304.594& 386.739\\
& $\quad $ 6 $\quad $ &1207.27 & 44.538 &47.587 &776.694&  396.079\\
& $\quad $ 9 $\quad $ & 1653.733 & 96.806 &57.775 & 1129.738& 424.694\\
& &&&&&\\
\multirow{3}{1em}{300}
& $\quad $ 3 $\quad $ &1868.786  & 26.726&71.125 & 802.264& 843.331\\
& $\quad $ 6 $\quad $ & 3233.860 &98.800 &103.029 &2030.358&  889.642\\
& $\quad $ 9 $\quad $ & 4417.411  & 199.993&125.012 & 3328.107& 947.897\\
\hline
\end{tabular}\label{tab:time}
\end{center}
\end{table}

\section{Discussion}
The design of computer experiments plays a crucial role in uncertainty quantification, surrogate modeling, and the development of robust Digital Twin systems. 
Space-filling designs, in particular, provide an efficient framework for exploring complex input-output relationships by ensuring uniform coverage of the experimental region. 
This review has examined key principles and methodologies in space-filling designs, including distance-based criteria (e.g., maximin and minimax designs), projection-based approaches (e.g., MaxPro and uniform projection designs), and advanced techniques for constrained or high-dimensional spaces.

Our discussion highlights that no single design is universally optimal—each has strengths depending on the problem context. 
Maximin designs excel in maximizing separation distance, while miniMax designs minimize the worst-case prediction error. 
Latin hypercube designs ensure one-dimensional uniformity, and their enhanced variants (e.g., orthogonal array-based or projection-optimized Latin hypercube designs) improve performance in lower-dimensional subspaces. 
Meanwhile, low-discrepancy sequences and non-uniform space-filling designs offer solutions for integration problems and constrained regions, respectively.

Emerging challenges—such as mixed variable types, multi-fidelity simulations, and high-dimensional spaces—demand further innovation. 
Recent developments, including grouped orthogonal arrays and sliced space-filling designs, demonstrate promising directions for accommodating complex experimental settings. Additionally, the integration of prior knowledge into design construction (e.g., through weighted or minimum energy designs) opens new avenues for improving efficiency in targeted applications.

Future research should focus on:
\begin{itemize}
\item Scalability--Developing computationally efficient algorithms for large-scale and high-dimensional problems.
\item Adaptivity--Incorporating sequential and active learning strategies to refine designs dynamically.
\item Generalizability--Extending space-filling principles to non-Euclidean and mixed input spaces (e.g., categorical and functional variables).
\item Integration with AI-- Leveraging machine learning to automate design selection and optimization.
\end{itemize}
As Digital Twin and simulation-based engineering continue to advance, the role of well-designed computer experiments will only grow in importance. 
By advancing both theoretical foundations and practical methodologies, this field can further enhance the reliability and predictive power of computational models across science and industry.

\section*{Acknowledgment}

The work of L. Kang was partially supported National Science Foundation grant DMS-2153029. 
Part of this research was performed while the three authors was visiting the
Institute for Mathematical and Statistical Innovation (IMSI) at University of Chicago from March 3 to May 24, 2025, which is supported by the National Science Foundation (Grant No. DMS-1929348).

\section*{Appendix A}

This section presents four test function based simulators used in Sections 4  for performance comparison of the competing design criteria for global fitting respectively. All test functions were scaled such that the inputs are in $[0,1]^d$.

\begin{itemize}

\item  the detpep10 function ($d=3$)  $$ 
	y(x) =  100\left(
    e^{-2/x_1^{1.75}} +e^{-2/x_2^{1.5}} + e^{-2/x_3^{1.25}}\right)
$$

\item  the Friedman function  ($d=5$)  
$$y(x) = 10\sin(\pi x_1x_2) + 20(x_3-0.5)^2 + 10x_4 + 5x_5.$$

\item  the GramacyLee function  ($d=6$)  
$$y(x) = e^{sin( (0.9(x_1+0.48))^{10} )} +x_2x_3+x_4$$

\item   the Bratley function  ($d=9$)  
$$y(x) = \sum_{i=1}^d (-1)^i \prod_{j=1}^ix_j.$$

\end{itemize}

\bibliographystyle{ECA_jasa}
\bibliography{reference} 

@article{sun2010construction,
  title={Construction of orthogonal {Latin} hypercube designs with flexible run sizes},
  author={Sun, F.\ and Liu, M.\ Q.\ and Lin, D.\ K.\ J.\ },
  journal={Journal of Statistical Planning and Inference},
  volume={140},
  number={11},
  pages={3236--3242},
  year={2010},
  publisher={Elsevier}
}

@article{georgiou2009orthogonal,
  title={Orthogonal {Latin} hypercube designs from generalized orthogonal designs},
  author={Georgiou, S.\ D.\ },
  journal={Journal of Statistical Planning and Inference},
  volume={139},
  number={4},
  pages={1530--1540},
  year={2009},
  publisher={Elsevier}
}

@article{yangconstruction,
  title={Construction of orthogonal and nearly orthogonal {Latin} hypercube designs from orthogonal designs},
  author={Yang, J.\ and Liu, M.\ Q.\ },
  journal={Statistica Sinica},
    volume={22},
  pages={433--442},
  year={2012}
}

@article{sun2009construction,
  title={Construction of orthogonal {Latin} hypercube designs},
  author={Sun, F.\ and Liu, M.\ Q.\ and Lin, D.\ K.\ J.\ },
  journal={Biometrika},
  volume={96},
  number={4},
  pages={971--974},
  year={2009},
  publisher={Biometrika Trust}
}

@misc{lin2008,
series = {Simon Fraser University. Theses (Department of Statistics and Actuarial Science)},
publisher = {Simon Fraser University},
year = {2008},
title = {New developments in designs for computer experiments and physical experiments / by Chunfang Devon Lin.},
language = {eng},
author = {Lin, Chunfang Devon},
}

@article{pang2009construction,
  title={A construction method for orthogonal {Latin} hypercube designs with prime power levels},
  author={Pang, F.\ and Liu, M.\ Q.\ and Lin, D.\ K.\ J.\ },
  journal={Statistica Sinica},
  volume={19},
  number={3},
    pages={1721--1728},
  year={2009}
}

@article{sun2017method,
  title={A method of constructing space-filling orthogonal designs},
  author={Sun, Fasheng and Tang, Boxin},
  journal={Journal of the American Statistical Association},
  volume={112},
  number={518},
  pages={683--689},
  year={2017},
  publisher={Taylor \& Francis}
}

@article{cioppa2007efficient,
  title={Efficient nearly orthogonal and space-filling {Latin} hypercubes},
  author={Cioppa, T.\ M.\ and Lucas, T.\ W.\ },
  journal={Technometrics},
  volume={49},
  number={1},
  pages={45--55},
  year={2007},
  publisher={ASA}
}

@article{ye1998orthogonal,
  title={Orthogonal column {Latin} hypercubes and their application in computer experiments},
  author={Ye, K.\ Q.\ },
  journal={Journal of the American Statistical Association},
    volume={93},
  pages={1430--1439},
  year={1998},
  publisher={JSTOR}
}

@article{tang1998selecting,
  title={Selecting {Latin} hypercubes using correlation criteria},
  author={Tang, B.\ },
  journal={Statistica Sinica},
  volume={8},
  pages={965--978},
  year={1998}
}

@article{box1959basis,
  title={A basis for the selection of a response surface design},
  author={Box, George EP and Draper, Norman R},
  journal={Journal of the American Statistical Association},
  volume={54},
  number={287},
  pages={622--654},
  year={1959},
  publisher={Taylor \& Francis}
}

@article{levy2010computer,
  title={Computer experiments: a review},
  author={Levy, Sigal and Steinberg, David M},
  journal={AStA Advances in Statistical Analysis},
  volume={94},
  pages={311--324},
  year={2010},
  publisher={Springer}
}

@article{Joseph02012016,
author = {V. Roshan Joseph},
title = {Space-filling designs for computer experiments: A review},
journal = {Quality Engineering},
volume = {28},
number = {1},
pages = {28--35},
year = {2016},
publisher = {Taylor \& Francis},
doi = {10.1080/08982112.2015.1100447}
}

@article{GARUD201771,
title = {Design of computer experiments: A review},
journal = {Computers \& Chemical Engineering},
volume = {106},
pages = {71-95},
year = {2017},
note = {ESCAPE-26},
issn = {0098-1354},
doi = {https://doi.org/10.1016/j.compchemeng.2017.05.010},
url = {https://www.sciencedirect.com/science/article/pii/S0098135417302090},
author = {Sushant S. Garud and Iftekhar A. Karimi and Markus Kraft}
}

@book{santner2003design,
  title={The design and analysis of computer experiments},
  author={Santner, Thomas J and Williams, Brian J and Notz, William I and Williams, Brain J},
  volume={1},
  year={2003},
  publisher={Springer},
  address={New York}
}

@article{kang2019stochastic,
  title={Stochastic coordinate-exchange optimal designs with complex constraints},
  author={Kang, Lulu},
  journal={Quality Engineering},
  volume={31},
  number={3},
  pages={401--416},
  year={2019},
  publisher={Taylor \& Francis}
}

@book{national2023foundational,
  title={Foundational research gaps and future directions for digital twins},
  author={{National Academies of Sciences, Engineering, and Medicine and others}},
  year={2023},
  publisher={The National Academies Press},
  address={Washington, DC},
  doi={https://doi.org/10.17226/26894}
}

@book{ghanem2017handbook,
  title={Handbook of uncertainty quantification},
  author={Ghanem, Roger and Higdon, David and Owhadi, Houman and others},
  volume={6},
  year={2017},
  publisher={Springer},
  address={New York}
}

@book{smith2024uncertainty,
author = {Smith, Ralph C.},
title = {Uncertainty Quantification: Theory, Implementation, and Applications, Second Edition},
publisher = {Society for Industrial and Applied Mathematics},
year = {2024},
doi = {10.1137/1.9781611977844},
address = {Philadelphia, PA},
edition   = {},
eprint = {https://epubs.siam.org/doi/pdf/10.1137/1.9781611977844}
}

@misc{QMCPy2020a,
  Author = {S.-C. T. Choi and F. J. Hickernell and M. McCourt and A. Sorokin},
  Title = {{QMCPy}: A quasi-{M}onte {C}arlo {P}ython Library},
  Year = {2020+}
}

@article{bratley1992implementation,
  title={Implementation and tests of low-discrepancy sequences},
  author={Bratley, Paul and Fox, Bennett L and Niederreiter, Harald},
  journal={ACM Transactions on Modeling and Computer Simulation (TOMACS)},
  volume={2},
  number={3},
  pages={195--213},
  year={1992},
  publisher={ACM New York, NY, USA}
}

@article{friedman1991multivariate,
  title={Multivariate adaptive regression splines},
  author={Friedman, Jerome H},
  journal={The Annals of Statistics},
  volume={19},
  number={1},
  pages={1--67},
  year={1991},
  publisher={Institute of Mathematical Statistics}
}

@article{dette2010generalized,
  title={Generalized {Latin} hypercube design for computer experiments},
  author={Dette, Holger and Pepelyshev, Andrey},
  journal={Technometrics},
  volume={52},
  number={4},
  pages={421--429},
  year={2010},
  publisher={Taylor \& Francis}
}

@article{gramacy2009adaptive,
  title={Adaptive design and analysis of supercomputer experiments},
  author={Gramacy, Robert B and Lee, Herbert KH},
  journal={Technometrics},
  volume={51},
  number={2},
  pages={130--145},
  year={2009},
  publisher={Taylor \& Francis}
}

@article{zhang2021distance,
  title={Distance-distributed design for Gaussian process surrogates},
  author={Zhang, Boya and Cole, D Austin and Gramacy, Robert B},
  journal={Technometrics},
  volume={63},
  number={1},
  pages={40--52},
  year={2021},
  publisher={Taylor \& Francis}
}

@article{he2024efficient,
author = {He, Xu},
title = {Efficient kriging using interleaved lattice-based designs with low fill and high separation distance properties},
journal = {SIAM/ASA Journal on Uncertainty Quantification},
volume = {12},
number = {4},
pages = {1113-1134},
year = {2024},
doi = {10.1137/23M156940X},
eprint = {https://doi.org/10.1137/23M156940X}
}

@article{tuo2020kriging,
  author  = {Rui Tuo and Wenjia Wang},
  title   = {Kriging Prediction with Isotropic Matern Correlations: Robustness and Experimental Designs},
  journal = {Journal of Machine Learning Research},
  year    = {2020},
  volume  = {21},
  number  = {187},
  pages   = {1--38}
}

@article{chen2025construction,
  title={Grouped orthogonal arrays and their construction method},
  author={Chen, Guanzhou and He, Yuanzhen and Lin, C Devon and Sun, Fasheng},
  journal={Statistica Sinica},
  volume={Accepted},
  year={2025} 
}

@article{he2017construction,
  title={On construction of marginally coupled designs},
  author={He, Yuanzhen and Lin, C Devon and Sun, Fasheng},
  journal={Statistica Sinica},
  volume  = {27},
  number  = {2},
  pages={665--683},
  year={2017},
  publisher={JSTOR}
}

@article{yang2023doubly,
  title={Doubly coupled designs for computer experiments with both qualitative and quantitative factors},
  author={Yang, Feng and Lin, C Devon and Zhou, Yongdao and He, Yuanzhen},
  journal={Statistica Sinica},
  volume={33},
  number={3},
  pages={1923--1942},
  year={2023},
  publisher={JSTOR}
}

@article{deng2015design,
  title={Design for computer experiments with qualitative and quantitative factors},
  author={Deng, Xinwei and Hung, Ying and Lin, C Devon},
  journal={Statistica Sinica},
  volume={25},
  pages={1567--1581},
  year={2015},
  publisher={JSTOR}
}

@article{deng2017additive,
  title={Additive Gaussian process for computer models with qualitative and quantitative factors},
  author={Deng, Xinwei and Lin, C Devon and Liu, K-W and Rowe, R Kerry},
  journal={Technometrics},
  volume={59},
  number={3},
  pages={283--292},
  year={2017},
  publisher={Taylor \& Francis}
}

@article{qian2008gaussian,
  title={Gaussian process models for computer experiments with qualitative and quantitative factors},
  author={Qian, Peter Z G and Wu, Huaiqing and Wu, CF Jeff},
  journal={Technometrics},
  volume={50},
  number={3},
  pages={383--396},
  year={2008},
  publisher={Taylor \& Francis}
}

@article{goh2013prediction,
  title={Prediction and computer model calibration using outputs from multifidelity simulators},
  author={Goh, Joslin and Bingham, Derek and Holloway, James Paul and Grosskopf, Michael J and Kuranz, Carolyn C and Rutter, Erica},
  journal={Technometrics},
  volume={55},
  number={4},
  pages={501--512},
  year={2013},
  publisher={Taylor \& Francis}
}

@article{yang2013construction,
  title={Construction of sliced orthogonal {Latin} hypercube designs},
  author={Yang, Jian-Feng and Lin, C Devon and Qian, Peter ZG and Lin, Dennis KJ},
  journal={Statistica Sinica},
  volume={26},
  number={2},
  pages={1117--1130},
  year={2013},
  publisher={JSTOR}
}

@article{qian2009sliced,
  title={Sliced space-filling designs},
  author={Qian, Peter ZG and Wu, CF Jeff},
  journal={Biometrika},
  volume={96},
  number={4},
  pages={945--956},
  year={2009},
  publisher={Oxford University Press}
}

@article{qian2009nested,
  title={Nested space-filling designs for computer experiments with two levels of accuracy},
  author={Qian, Peter ZG and Tang, Boxin and Wu, CF Jeff},
  journal={Statistica Sinica},
  volume={19},
  number={1},
  pages={287--300},
  year={2009},
  publisher={JSTOR}
}

@article{chen2024selecting,
  title={Selecting strong orthogonal arrays by linear allowable level permutations},
  author={Chen, Guanzhou and Tang, Boxin},
  journal={Electronic Journal of Statistics},
  volume={18},
  number={2},
  pages={3573--3589},
  year={2024},
  publisher={The Institute of Mathematical Statistics and the Bernoulli Society}
}

@article{he2013strong,
  title={Strong orthogonal arrays and associated {Latin} hypercubes for computer experiments},
  author={He, Yuanzhen and Tang, Boxin},
  journal={Biometrika},
  volume={100},
  number={1},
  pages={254--260},
  year={2013},
  publisher={Oxford University Press}
}

@article{hickernell1998generalized,
  title={A generalized discrepancy and quadrature error bound},
  author={Hickernell, Fred},
  journal={Mathematics of Computation},
  volume={67},
  number={221},
  pages={299--322},
  year={1998}
}

@article{joseph2015sequential,
  title={Sequential exploration of complex surfaces using minimum energy designs},
  author={Joseph, V Roshan and Dasgupta, Tirthankar and Tuo, Rui and Wu, CF Jeff},
  journal={Technometrics},
  volume={57},
  number={1},
  pages={64--74},
  year={2015},
  publisher={Taylor \& Francis}
}

@article{lu2021non,
  title={Non-uniform space filling (NUSF) designs},
  author={Lu, Lu and Anderson-Cook, Christine M and Ahmed, Towfiq},
  journal={Journal of Quality Technology},
  volume={53},
  number={3},
  pages={309--330},
  year={2021},
  publisher={Taylor \& Francis}
}

@article{petelet2010latin,
  title={Latin hypercube sampling with inequality constraints},
  author={Petelet, Matthieu and Iooss, Bertrand and Asserin, Olivier and Loredo, Alexandre},
  journal={AStA Advances in Statistical Analysis},
  volume={94},
  pages={325--339},
  year={2010},
  publisher={Springer}
}

@article{stinstra2003constrained,
  title={Constrained maximin designs for computer experiments},
  author={Stinstra, Erwin and den Hertog, Dick and Stehouwer, Peter and Vestjens, Arjen},
  journal={Technometrics},
  volume={45},
  number={4},
  pages={340--346},
  year={2003},
  publisher={Taylor \& Francis}
}

@article{iman1982distribution,
  title={A distribution-free approach to inducing rank correlation among input variables},
  author={Iman, Ronald L and Conover, William-Jay},
  journal={Communications in Statistics-Simulation and Computation},
  volume={11},
  number={3},
  pages={311--334},
  year={1982},
  publisher={Taylor \& Francis}
}

@article{bowman2013weighted,
  title={Weighted space-filling designs},
  author={Bowman, Veronica E and Woods, David C},
  journal={Journal of Simulation},
  volume={7},
  number={4},
  pages={249--263},
  year={2013},
  publisher={Taylor \& Francis}
}

@article{chen2019optimal,
  title={Optimal noncollapsing space-filling designs for irregular experimental regions},
  author={Chen, Ray-Bing and Li, Chi-Hao and Hung, Ying and Wang, Weichung},
  journal={Journal of Computational and Graphical Statistics},
  volume={28},
  number={1},
  pages={74--91},
  year={2019},
  publisher={Taylor \& Francis}
}

@article{lekivetz2015fast,
  title={Fast flexible space-filling designs for nonrectangular regions},
  author={Lekivetz, Ryan and Jones, Bradley},
  journal={Quality and Reliability Engineering International},
  volume={31},
  number={5},
  pages={829--837},
  year={2015},
  publisher={Wiley Online Library}
}

@article{draguljic2012noncollapsing,
  title={Noncollapsing space-filling designs for bounded nonrectangular regions},
  author={Dragulji{\'c}, Danel and Santner, Thomas J and Dean, Angela M},
  journal={Technometrics},
  volume={54},
  number={2},
  pages={169--178},
  year={2012},
  publisher={Taylor \& Francis}
}

@book{niederreiter1992random,
  title={Random number generation and quasi-Monte Carlo methods},
  author={Niederreiter, Harald},
  year={1992},
  publisher={SIAM},
  address={Philadelphia, Pennsylvania}
}

@article{halton1960efficiency,
  title={On the efficiency of certain quasi-random sequences of points in evaluating multi-dimensional integrals},
  author={Halton, John H},
  journal={Numerische Mathematik},
  volume={2},
  pages={84--90},
  year={1960},
  publisher={Springer}
}

@article{sobol1967distribution,
  title={The distribution of points in a cube and the approximate evaluation of integrals},
  author={Sobol, Ilya M},
  journal={USSR Computational Mathematics and Mathematical Physics},
  volume={7},
  pages={86--112},
  year={1967}
}

@article{georgiou2014construction,
  title={Construction of orthogonal and nearly orthogonal designs for computer experiments},
  author={Georgiou, SD and Stylianou, Stella and Drosou, K and Koukouvinos, C},
  journal={Biometrika},
  volume={101},
  number={3},
  pages={741--747},
  year={2014},
  publisher={Oxford University Press}
}

@article{bingham2009orthogonal,
  title={Orthogonal and nearly orthogonal designs for computer experiments},
  author={Bingham, Derek and Sitter, Randy R and Tang, Boxin},
  journal={Biometrika},
  volume={96},
  number={1},
  pages={51--65},
  year={2009},
  publisher={Oxford University Press}
}

@Manual{maximin,
    title = {maximin: Space-Filling Design under Maximin Distance},
    author = {Furong Sun and Robert B. Gramacy },
    year = {2024},
  }

@Manual{SLHD,
    title = {SLHD: Maximin-Distance (Sliced) {Latin} Hypercube Designs},
    author = {Shan Ba},
    year = {2015},
  }

@Manual{DiceDesign,
    title = {DiceDesign: Designs of Computer Experiments},
    author = {	Jessica Franco and Delphine Dupuy and Olivier Roustan abd  Patrice Kiener and Guillaume Damblin and Bertrand Iooss},
    year = {2025},
  }

@Manual{DiceKriging,
    title = {DiceKriging: Kriging Methods for Computer Experiments},
    author = {Olivier Roustant and David Ginsbourger and Yves Deville},
    year = {2021},
  }

@article{sun2017general,
  title={A general rotation method for orthogonal {Latin} hypercubes},
  author={Sun, Fasheng and Tang, Boxin},
  journal={Biometrika},
  volume={104},
  number={2},
  pages={465--472},
  year={2017},
  publisher={Oxford University Press}
}

@article{lin2010new,
  title={A new and flexible method for constructing designs for computer experiments},
  author={Lin, C Devon and Bingham, Derek and Sitter, Randy R and Tang, Boxin},
  journal={The Annals of Statistics},
  volume={38},
  number={3},
  pages={1460--1477},
  year={2010},
  publisher={JSTOR}
}

@article{lin2009construction,
  title={Construction of orthogonal and nearly orthogonal {Latin} hypercubes},
  author={Lin, C Devon and Mukerjee, Rahul and Tang, Boxin},
  journal={Biometrika},
  volume={96},
  number={1},
  pages={243--247},
  year={2009},
  publisher={Oxford University Press}
}

@article{steinberg2006construction,
  title={A construction method for orthogonal {Latin} hypercube designs},
  author={Steinberg, David M and Lin, Dennis KJ},
  journal={Biometrika},
  volume={93},
  number={2},
  pages={279--288},
  year={2006},
  publisher={Oxford University Press}
}

@article{owen1994controlling,
  title={Controlling correlations in {Latin} hypercube samples},
  author={Owen, Art B},
  journal={Journal of the American Statistical Association},
  volume={89},
  number={428},
  pages={1517--1522},
  year={1994},
  publisher={Taylor \& Francis}
}

@article{owen1992orthogonal,
  title={Orthogonal arrays for computer experiments, integration and visualization},
  author={Owen, Art B},
  journal={Statistica Sinica},
  volume={2},
  number={2},
  pages={439--452},
  year={1992},
  publisher={JSTOR}
}

@article{tang1993orthogonal,
  title={Orthogonal array-based {Latin} hypercubes},
  author={Tang, Boxin},
  journal={Journal of the American statistical association},
  volume={88},
  number={424},
  pages={1392--1397},
  year={1993},
  publisher={Taylor \& Francis}
}

@article{zhou2015space,
  title={Space-filling properties of good lattice point sets},
  author={Zhou, Yongdao and Xu, Hongquan},
  journal={Biometrika},
  volume={102},
  number={4},
  pages={959--966},
  year={2015},
  publisher={Oxford University Press}
}

@article{yuan2025construction,
  title={A construction method for maximin L1-distance {Latin} hypercube designs},
  author={Yuan, Ru and Yin, Yuhao and Xu, Hongquan and Liu, Min-Qian},
  journal={Statistica Sinica},
  volume={35},
  pages={249--272},
  year={2025}
}

@article{lin2016general,
  title={A general construction for space-filling {Latin} hypercubes},
  author={Lin, C Devon and Kang, L},
  journal={Statistica Sinica},
  volume={26},
  number={2},
  pages={675--690},
  year={2016},
  publisher={JSTOR}
}

@article{van2008two,
  title={Two-dimensional minimax {Latin} hypercube designs},
  author={van Dam, Edwin R},
  journal={Discrete Applied Mathematics},
  volume={156},
  number={18},
  pages={3483--3493},
  year={2008},
  publisher={Elsevier}
}

@article{mak2018minimax,
  title={Minimax and minimax projection designs using clustering},
  author={Mak, Simon and Joseph, V Roshan},
  journal={Journal of Computational and Graphical Statistics},
  volume={27},
  number={1},
  pages={166--178},
  year={2018},
  publisher={Taylor \& Francis}
}

@article{he2017interleaved,
  title={Interleaved lattice-based minimax distance designs},
  author={He, Xu},
  journal={Biometrika},
  volume={104},
  number={3},
  pages={713--725},
  year={2017},
  publisher={Oxford University Press}
}

@article{sun2019uniform,
  title={Uniform projection designs},
  author={Sun, Fasheng and Wang, Yaping and Xu, Hongquan},
  journal={The Annals of Statistics},
  volume={47},
  number={1},
  pages={641--661},
  year={2019}
}

@article{ba2015optimal,
  title={Optimal sliced {Latin} hypercube designs},
  author={Ba, Shan and Myers, William R and Brenneman, William A},
  journal={Technometrics},
  volume={57},
  number={4},
  pages={479--487},
  year={2015},
  publisher={Taylor \& Francis}
}

@article{yin2023construction,
  title={Construction of maximin $L_1$-distance {Latin} hypercube designs},
  author={Yin, Yuhao and Wang, Lin and Xu, Hongquan},
  journal={Electronic Journal of Statistics},
  volume={17},
  number={2},
  pages={3942--3968},
  year={2023},
  publisher={The Institute of Mathematical Statistics and the Bernoulli Society}
}

@article{xiao2017construction,
  title={Construction of maximin distance {Latin} squares and related {Latin} hypercube designs},
  author={Xiao, Qian and Xu, Hongquan},
  journal={Biometrika},
  volume={104},
  number={2},
  pages={455--464},
  year={2017},
  publisher={Oxford University Press}
}

@article{li2021method,
  title={A method of constructing maximin distance designs},
  author={Li, Wenlong and Liu, Min-Qian and Tang, Boxin},
  journal={Biometrika},
  volume={108},
  number={4},
  pages={845--855},
  year={2021},
  publisher={Oxford University Press}
}

@article{lin2015latin,
  title={Latin hypercubes and space-filling designs},
  author={Lin, C Devon and Tang, Boxin},
  journal={Handbook of Design and Analysis of Experiments},
  pages={593--625},
  year={2015},
  publisher={Chapman and Hall/CRC New York}
}

@article{wang2022design,
  title={On design orthogonality, maximin distance, and projection uniformity for computer experiments},
  author={Wang, Yaping and Sun, Fasheng and Xu, Hongquan},
  journal={Journal of the American Statistical Association},
  volume={117},
  number={537},
  pages={375--385},
  year={2022},
  publisher={Taylor \& Francis}
}

@article{van2007maximin,
  title={Maximin {Latin} hypercube designs in two dimensions},
  author={Van Dam, Edwin R and Husslage, Bart and Den Hertog, Dick and Melissen, Hans},
  journal={Operations Research},
  volume={55},
  number={1},
  pages={158--169},
  year={2007},
  publisher={INFORMS}
}

@article{shewry1987maximum,
  title={Maximum entropy sampling},
  author={Shewry, Michael C and Wynn, Henry P},
  journal={Journal of Applied Statistics},
  volume={14},
  number={2},
  pages={165--170},
  year={1987},
  publisher={Taylor \& Francis}
}

@article{MoonDeanSantner2011,
  title={Algorithms for generating maximin orthogonal and {Latin} hypercube designs},
  author={Moon, H.\ and Dean, A.\ M.\ and Santner, T.\ J.\ },
  journal={Journal of Statistical Theory and Practice},
    volume={5},
  pages={81--98},
  year={2011},
  publisher={JSTOR}
}

@article{audze1977new,
  title={New approach to planning out of experiments},
  author={Audze, P.\ and Eglais, V.\ },
  journal={Problems of Dynamics and Strength},
  volume={35},
  pages={104--107},
  year={1977}
}

@article{loh1996latin,
  title={On Latin hypercube sampling},
  author={Loh, Wei-Liem},
  journal={The Annals of Statistics},
  volume={24},
  number={5},
  pages={2058--2080},
  year={1996},
  publisher={Institute of Mathematical Statistics}
}

@article{owen1992central,
  title={A central limit theorem for {Latin} hypercube sampling},
  author={Owen, A.\ B.\ },
  journal={Journal of the Royal Statistical Society. Series B (Methodological)},
    volume={54},
  pages={541--551},
  year={1992},
  publisher={JSTOR}
}

@article{stein1987large,
  title={Large sample properties of simulations using {Latin} hypercube sampling},
  author={Stein, M.\ },
  journal={Technometrics},
    volume={29},
  pages={143--151},
  year={1987},
  publisher={JSTOR}
}

@article{patterson1954errors,
  title={The errors of lattice sampling},
  author={Patterson, H.\ D.\ },
  journal={Journal of the Royal Statistical Society. Series B (Methodological)},
    volume={16},
  pages={140--149},
  year={1954},
  publisher={JSTOR}
}

@article{mckay1979comparison,
  title={A comparison of three methods for selecting values of input variables in the analysis of output from a computer code},
  author={McKay, Michael D and Beckman, Richard J and Conover, William J},
  journal={Technometrics},
  volume={21},
  pages={239--245},
  year={1979},
  publisher={Taylor \& Francis}
}

@article{pronzato2017minimax,
  title={Minimax and maximin space-filling designs: some properties and methods for construction},
  author={Pronzato, Luc},
  journal={Journal de la Soci{\'e}t{\'e} Fran{\c{c}}aise de Statistique},
  volume={158},
  number={1},
  pages={7--36},
  year={2017}
}

@article{johnson1990minimax,
  title={Minimax and maximin distance designs},
  author={Johnson, Mark E and Moore, Leslie M and Ylvisaker, Donald},
  journal={Journal of Statistical Planning and Inference},
  volume={26},
  number={2},
  pages={131--148},
  year={1990},
  publisher={Elsevier}
}

@book{wendland2004scattered,
  title={Scattered Data Approximation},
  author={Wendland, Holger},
  volume={17},
  year={2004},
  publisher={Cambridge University Press},
  address={Cambridge}
}

@article{pronzato2012design,
  title={Design of computer experiments: space-filling and beyond},
  author={Pronzato, Luc and M{\"u}ller, Werner G},
  journal={Statistics and Computing},
  volume={22},
  pages={681--701},
  year={2012},
  publisher={Springer}
}

@article{peng2014on,
author = {Chien-Yu Peng and C. F. Jeff Wu},
title = {On the choice of nugget in Kriging modeling for deterministic computer experiments},
journal = {Journal of Computational and Graphical Statistics},
volume = {23},
number = {1},
pages = {151-168},
year  = {2014},
publisher = {Taylor \& Francis},
doi = {10.1080/10618600.2012.738961},
eprint = {https://doi.org/10.1080/10618600.2012.738961}
}

@article{joseph2008orthogonal,
  title={Orthogonal-maximin Latin hypercube designs},
  author={Joseph, V Roshan and Hung, Ying},
  journal={Statistica Sinica},
  pages={171--186},
  year={2008},
  publisher={JSTOR}
}

@book{fang2002,
  author = {Fang, K.-T. and Li, R. and Sudjianto, A.},
  title = {Design and Modeling for Computer Experiments},
  publisher = {Chapman \& Hall/CRC},
  address={New York},
  year = {2002}
}

@article{morris1995,
  author = {Morris, M. D. and Mitchell, T. J.},
  title = {Exploratory designs for computational experiments},
  journal = {Journal of Statistical Planning and Inference},
  volume = {43},
  number = {3},
  pages = {381-402},
  year = {1995}
}

@article{joseph2015,
  author = {Joseph, V. R. and Gul, E. and Ba, S.},
  title = {Maximum projection designs for computer experiments},
  journal = {Biometrika},
  volume = {102},
  number = {2},
  pages = {371-380},
  year = {2015}
}

\end{document}